\documentclass[12pt,preprint]{aastex}

\newcommand{\OIII}{[O{\footnotesize III}]$\lambda$5007}  
\newcommand{\Ha}{H$\alpha$}  
\newcommand{\Hb}{H$\beta$}

\newcommand{\Msolar}{M$_{\odot}$}                           
  
\shorttitle{ Thermal Emission from  HII Galaxies }
\shortauthors{Rosa-Gonz\'alez et al. }

\begin{document}


\title{ Thermal Emission from  HII Galaxies: \\ Discovering the Youngest Systems}


\author{D. Rosa-Gonz\'alez}
\affil{ Instituto Nacional de Astrof\'{\i}sica Optica y Electr\'onica.
Luis Enrique Erro No. 1. Tonantzintla, Puebla, C.P. 72840, M\'exico; 
danrosa@inaoep.mx.} 
\author{H. R. Schmitt}
\affil{Remote Sensing Division, Code 7210, Naval Research Laboratory,
4555 Overlook Avenue, Washington, DC\,20375, USA; and\\
Interferometrics, Inc., 13454 Sunrise Valley Drive, Suite 240, Herndon,
VA\,20171, USA; henrique.schmitt@nrl.navy.mil.}

\and

\author{E. Terlevich\altaffilmark{1} and R. Terlevich\altaffilmark{1}}
\affil{Instituto Nacional de Astrof\'{\i}sica Optica y Electr\'onica.
Luis Enrique Erro No. 1. Tonantzintla, Puebla, C.P. 72840, M\'exico; 
eterlevi@inaoep.mx, rjt@inaoep.mx.
}


\altaffiltext{1}{Visiting Fellow, IoA, Cambridge. }


\begin{abstract}
We studied the radio properties of very young massive regions of star 
formation in HII galaxies, with the aim of detecting episodes of recent 
star formation in an early phase of evolution where the first supernovae
start to appear. Our sample consists of 31 HII galaxies, characterized by
strong Hydrogen emission lines, for which low resolution VLA 3.5~cm and
6~cm observations were obtained. We complemented these observations with
archival data at 20~cm. We found that the radio spectral energy distribution
(SED) has a range of behaviours; 1) there are  galaxies where the SED
is characterized by a synchrotron-type slope, 2) galaxies with a thermal 
slope, and, 3) galaxies with possible free-free absorption at long wavelengths.
The latter SEDs were found in a few galaxies (e.g. UM\,533, Tololo~1223-388)
and represent a signature of heavily embedded massive star clusters
closely related to the early stages of massive star formation.

Based on the comparison of the star formation rates (SFR) determined from
the recombination lines and those determined from the radio emission we find
that SFR(H$\alpha$) is on average five times higher than SFR(1.4~GHz).
We confirm this tendency by comparing the ratio between the observed
flux at 20 cm and the expected one, calculated based on the
H$\alpha$ star formation rates, both for the galaxies in our sample 
and for 
normal ones. This analysis shows that this ratio is a factor of 2 smaller
in our galaxies than in normal ones, indicating that they fall below the 
FIR/radio correlation. This result is further confirmed by the detection
of high q-parameter values (the ratio of infrared to radio fluxes) in a few
sources. These results suggest that the emission of these
galaxies is dominated by a recent and massive star formation event
in which the first supernovae (SN) just started to explode. This indicates
that the radio emission is most likely dominated by free-free continuum, and
that the emission at low frequencies may be optically thick, in agreement
with the observed SEDs.

We combined the VLA data with age indicators based on optical observations
(e.g.~equivalent width of \Hb) together with the ratio between the far
infrared and the radio continuum fluxes and proposed an evolutionary scenario
to explain the observed trends.  We conclude that the systematic lack of 
synchrotron emission in those systems with the largest equivalent width of \Hb\
can only be explained if those are young starbursts 
 of less than 3.5Myr of age, i.e.~before the first type II SNe  
start to explode.

\end{abstract}




\section{Introduction}

One of the outstanding problems in the study of starburst galaxies is 
understanding how massive stellar clusters (MSC) form and evolve. 
Most of what is known about MSC  comes from the study of relatively 
old objects, since they evolve
fast and can be heavily embedded in their earlier stages, thus being
difficult to detect (Johnson 2004). Recent attempts using radio observations
have been successful in the detection of several of these embedded sources in nearby 
starburst galaxies. During the early stages of evolution, very young star
clusters should either lack or have a deficit of synchrotron emission, being dominated
by free-free radio emission. Given that the synchrotron emission observed in
starbursts is due to supernova activity, it is a direct tracer of the end
of the evolution of massive stars (M\,$\ge$ 8\,M$_{\odot}$). In the case of
a coeval star formation burst, the first type II SNe appear
after only 3.5\,Myr and last until the 8\,M$_{\odot}$ stars explode at around
40\,Myr. Consequently, the lack of synchrotron emission is a good indicator of
very young clusters.
 
Recent radio observations (e.g. Turner, Beck, \& Ho 2000; 
Johnson et al. 2001; Johnson et al. 2003; Cannon \& Skillman 2004) 
have detected in a few nearby starburst galaxies, sources with compact 
flat radio spectrum which can be securely identified as
young star clusters. In some of the cases these authors detect an inverted
spectrum at low frequencies, due to free-free absorption, indicating that
the cluster still is deeply embedded in the material from which it was formed.
On a more global scale, Roussel et al. (2003) detected galaxies with
abnormally high infrared to radio continuum ratios, identified as nascent
starbursts. 

One important question is how MSC do form? Are they the product of an 
instantaneous
event or the star formation activity lasts several Myr instead? The 
situation is not clear,
the excellent radio-FIR correlation (e.g. Yun et al. 2001) 
strongly suggests that star formation,
at least at galactic scales, must proceed continuously; also in a study
of the statistical properties of HII galaxies we found strong 
indications of continuous star formation (Terlevich et al.~2004). 
On the other hand, the presence
of strong WR features in the spectra of starbursts and HII galaxies 
suggests a high degree of synchronization in the formation of massive stars.
The question may be solved studying the properties of a complete sample of
young systems.

Although these young starbursts are known, their numbers are
relatively small. The detection of a larger number of young sources is an
important step in the study of the early stages of massive star formation
and evolution. This will allow us to better understand their properties
and environment.

For this work we have taken a different approach, which consists in
determining the radio properties of a sample of HII galaxies selected
because their emission line spectra indicate extreme youth.
HII galaxies are low mass objects whose emission and thus most observables
are dominated by a massive burst of star formation. Their
optical spectrum is identical to that observed in giant HII regions
like 30-Doradus in the LMC. These properties make these galaxies ideal
targets for a study of integrated properties in search for features
related to MSCs. The best sources can then be studied in more detail with
higher resolution observations.
In this sense HII galaxies are considered ``young". In fact they are probably
the youngest stellar systems that can be studied in any detail. The youth 
scenario is supported not only by the strong emission lines from ionized 
gas, but also by their underlying stellar continuum properties, i.e. no stellar
absorptions are detected in those HII galaxies with the strongest emission 
lines (EW(H$\beta) > $150\AA) and only weak hydrogen and helium and some very
weak metal lines are detected on those more evolved HII galaxies
(EW(H$\beta) < $ 50\AA).
The lack of detection of stellar absorptions in the extreme HII galaxies
is due to the fact that the optical spectrum of very young clusters is 
dominated
by the light of massive blue supergiants close to the main sequence 
turn-off.
The spectrum of these stars shows narrow and relatively weak absorptions
of hydrogen and helium plus extremely weak metal lines. In the case
of HII galaxies (and also HII regions) the weak and narrow stellar 
absorptions
are filled with strong emission from the ionized gas of the associated HII 
region
making it impossible to  detect directly their presence. 
This has been traditionally a problem for studies to derive highly precise
helium abundances  as required for example for primordial 
helium determinations 
(e.g.~Skillman, Terlevich \& Terlevich 1998).
Thus very strong emission plus a ``featureless'' continuum is the signature 
of extremely young HII galaxies.

This ``youth" of the HII galaxies does not imply that the whole galaxy is 
young, in fact
the opposite is probably true given the colors of the extensions detected 
in some HII galaxies (Telles \& Terlevich, 1997) 
and the simple fact that the metal content of all HII galaxies 
is typically around 0.1-0.2 of solar (e.g.~Hoyos \& D\'\i az 2006)
implying a substantial amount of 
chemical evolution or astration previous to the present dominant burst.
Stellar population synthesis analysis
(Raimann et al. 2000a, Raimann et al. 2000b) show that there is some
evidence
for older stellar populations in these sources, albeit at a very low level,
again
confirming that their emission is in fact dominated by recent star
formation.

Thus ``young" and ``age" in this context refers to the present burst.
HII galaxies could be, and probably are, much older.

Here we present the results of a study of integrated radio spectral energy 
distributions (SEDs) of
nascent starburst candidates. We find several galaxies with flat spectrum,
typical of thermal emission, or even with inverted spectra, indicating
the presence of heavily embedded star clusters. With the data at hand we
compare the radio and emission line (H$\alpha$,H$\beta$) properties of these
galaxies and propose a qualitative evolutionary model. 

This paper is organized in the following way. In Section~\ref{Sec:Sample} we
describe how the sample was selected. Section~\ref{Sec:Observatios} 
describes the VLA observations at 6~cm (4.9~GHz) and 3.5~cm (8.4~GHz)
and the data reduction. Section~\ref{Sec:Radio} presents the radio SEDs,
spectral indices and the proposed evolutionary model. In 
Section~\ref{Sec:Opt} we study the optical properties of the sample
galaxies and calculate the SFRs obtained from the optical emission lines.
In Section~\ref{Sec:OpticalRadio}
we combined optical and radio data to discuss the nature of the selected 
galaxies and explore simple evolutionary scenarios.
In Section~\ref{Sec:FIR} we estimate the FIR to radio continuum flux ratios
($q$ parameter) and its relation to 
other observables to confirm the young nature of some of the observed 
galaxies. Finally, a summary of our results is presented in 
Section~\ref{Sec:Discussion}.

\section{Sample Selection}
\label{Sec:Sample}

Our sample selection was based on the expectation that in young starforming galaxies
their most massive  stars did not have enough time to evolve and explode as supernovae. In a
situation like this, these galaxies should have very little synchrotron emission
(Bressan, Silva \& Granato  2002).
Consequently, their radio luminosities should be smaller than what one
would expect to find based on the SFR measured from other indicators, like
H$\alpha$. 

For the current study we selected a sample from the catalog of HII
galaxies  by Terlevich et al. (1991; hereafter HIIG91). Based on the 
reddening corrected
H$\alpha$ fluxes of the galaxies in this catalog, we estimate their Star
Formation Rates (SFR) using the relation given by Rosa-Gonz\'alez, Terlevich
and Terlevich (2002). These SFRs were then converted into radio 1.4~GHz 
fluxes,
using the relation given by Schmitt et al. (2006). Comparing the predicted
1.4~GHz fluxes with values obtained from the NRAO/VLA Sky Survey 
(NVSS) catalog, we selected those
galaxies which were not detected at the NVSS 5$\sigma$ limit (2.5~mJy), or have
an observed flux smaller than the predicted one.

Our final sample is composed of 31 galaxies which are presented in Table~\ref{Tab:RadioReconst}.
Details about the VLA observations are described below.

\section{Observations}
\label{Sec:Observatios}

Our observations were obtained with the VLA in 2004-Sep-01, in the early 
stages
of the reconfiguration from D to A array. The galaxies were observed at
4.9~GHz (6~cm) and 8.4~GHz (3.5~cm), in snapshot mode, integrating for
$\sim$10~minutes per frequency. Observations of the sources were immediately
preceded and followed by $\sim$2~minutes observations of a phase calibrator.
We used calibrators from the NRAO list, preferentially closer than
10$^{\circ}$ from the galaxies. All the observations were done in
standard continuum mode with 2 IF's of 50~MHz bandwidth each, using the
radio galaxies 3C\,48 and 3C\,286 as primary calibrators. 
For the sources previously observed by the VLA we used archival data
to complement our observations.

The reductions followed standard AIPS techniques, which consisted of flagging
bad data points, setting the flux-density scale using the primary calibrators
and phase calibrating using the secondary calibrators. Since the
observations happened in the early stages of the reconfiguration, most of the
antennas were still in the D-configuration. Given that we are interested
only in the integrated properties of these galaxies, we flagged the 6 antennas
that had already been moved to the A configuration. Images were created using
uniform weighting. For a few sources we compared these images with ones
created using natural weighting, which is more sensitive to the extended
emission, but did not find any difference. This is due to the fact that we
flagged the outer antennas and were sensitive only to the large scale emission.
For those sources with peak flux densities of $\sim$1~mJy
or higher, we interactively self-calibrate them two or three times in phase. 

The noise of the images was determined in regions free from emission.
The total Stokes~{\bf \small I} fluxes were obtained by integrating the regions
brighter than 3$\sigma$ above the background level. All the sources
are unresolved by our observations.
The errors in the flux measurements were calculated by taking into
account, in quadrature, a 1.5\% uncertainty in the flux calibration
and Poisson noise, which usually dominates the errors. Table~1 gives the
galaxies observed, the VLA project from which they were obtained and the
image reconstruction beam parameters. The 1.4~GHz data 
were obtained from the NVSS catalog. For those galaxies
that were not detected at this frequency we list the reconstruction
beam of the nearest detectable source for reference.

\section{Radio Spectral Energy Distribution }
\label{Sec:Radio}

\begin{figure} 
\epsscale{1.0} 
\plotone{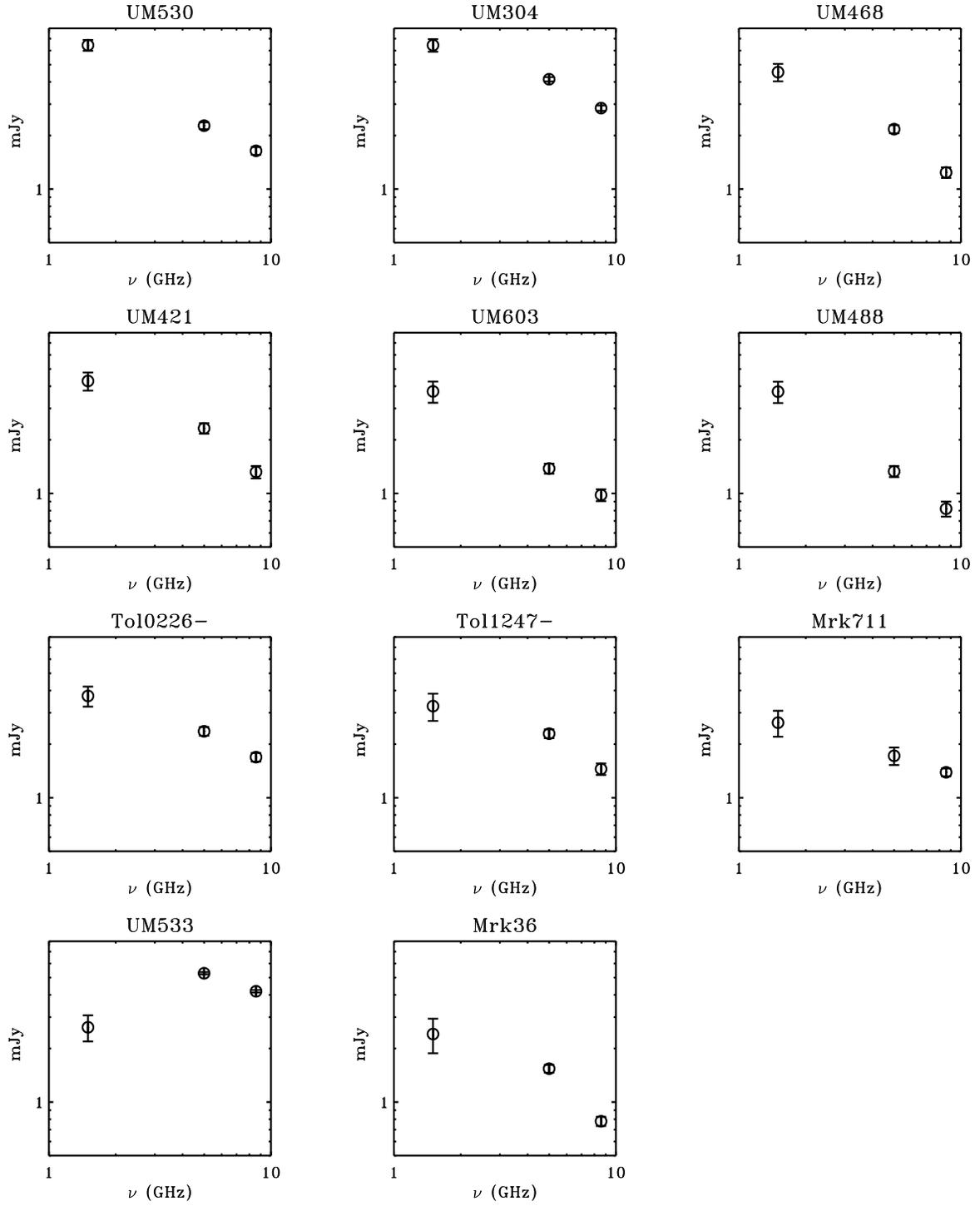} 
\caption{\label{Fig:SFR1} Radio SEDs of galaxies detected at all frequencies.}
\end{figure}

\begin{figure} 
\epsscale{1.0} 
\plotone{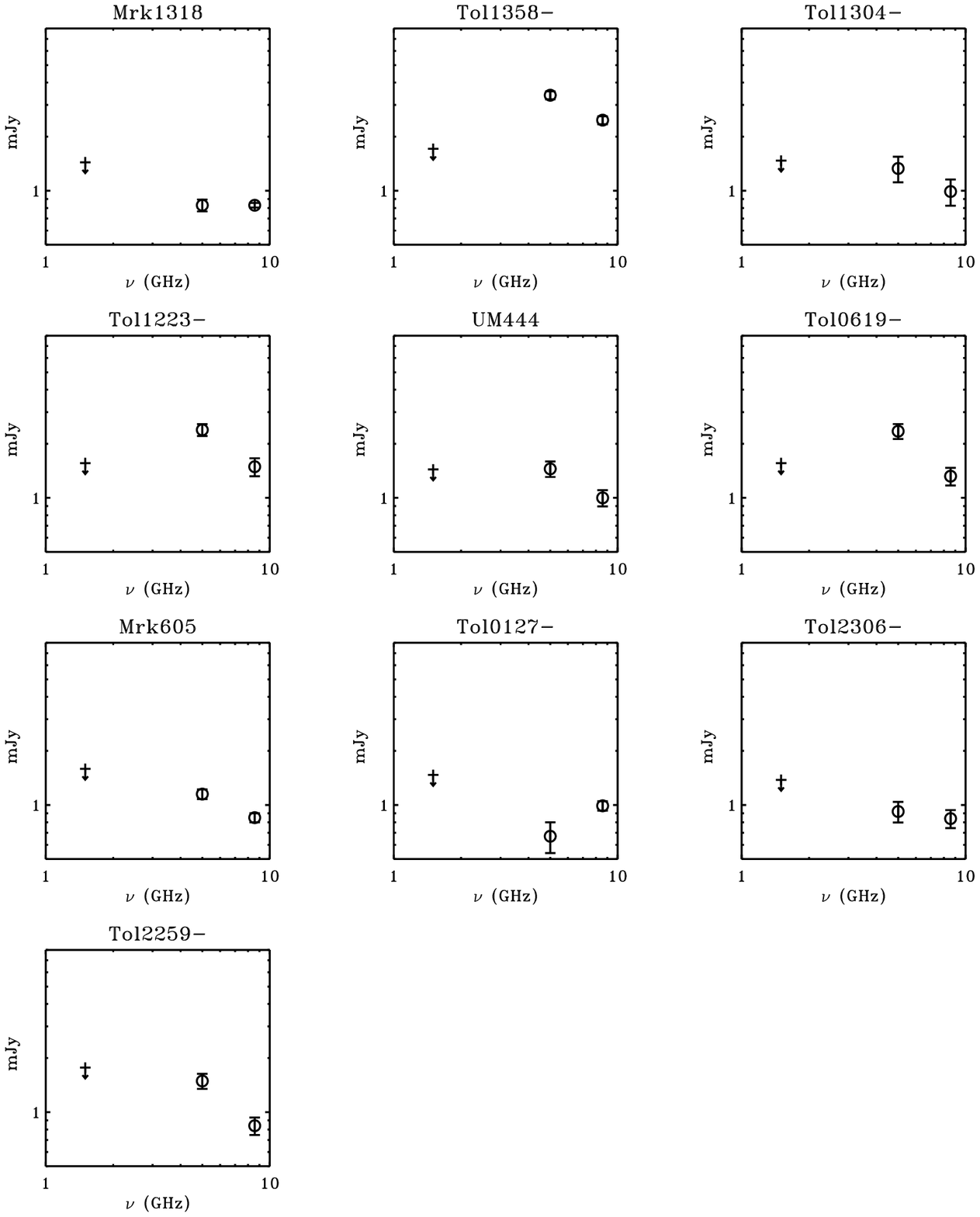} 
\caption{\label{Fig:SFR2} Same as Figure 1, but for galaxies not detected at
20~cm. The upper limits at 1.4~GHz correspond to the 3$\sigma$ value.} 
\end{figure}

In Figures~\ref{Fig:SFR1}~and~\ref{Fig:SFR2} we present the radio SEDs
of the observed galaxies. The fluxes used for these Figures are presented
in Table~\ref{Tab:RadioData}, where we also give the spectral indices 
($\alpha,F_\nu\propto \nu^\alpha$), 
between 1.4 and 4.9 GHz ($\alpha^{1.4}_{5}$) 
and between 4.9 and 8.4 GHz ($\alpha^{5}_{8}$), calculated using these fluxes.
Figure~\ref{Fig:SFR1} presents the galaxies that were detected at all
wavelengths, while Figure~\ref{Fig:SFR2} presents the sources for which
the 1.4~GHz flux is only an upper limit. In cases like this we calculated
an upper limit 
for $\alpha^{1.4}_{5}$ assuming a 1.4~GHz flux corresponding
to 3 times the r.m.s. value of the NVSS image. Notice also that for some
of the sources presented in Table~\ref{Tab:RadioData} (e.g. MRK\,930)
we do not give information about spectral indices, even though these
galaxies were detected in at least a couple of bands. This is due to
the fact that different frequencies were observed with significantly
different configurations. As a result, one of the frequencies was not
properly sampled at short baselines, thus missing the diffuse, more
extended emission. Calculating spectral indices with these data would
produce meaningless results.

Analyzing the spectral slopes of the galaxies detected at 1.4~GHz
we find that they have $\alpha^{1.4}_{5}$ values between those expected
for thermal emission ($\alpha=-0.1$), and those characteristic of
a source dominated by synchrotron radiation (typical value is around 
$\alpha=-0.8$, Condon 1992, spanning a range as can be seen from Figures 3 
and 6).
An interesting exception is UM\,533, which is the only galaxy
detected at all frequencies, with $\alpha^{1.4}_{5}>0$. If we consider
the galaxies that were not detected at 1.4~GHz we find other 4 sources
with spectral slopes in this range (Tol\,0619-392, UM\,444, Tol\,1223-388,
and Tol\,1358-328). These galaxies are of particular interest for our study,
because these spectral indices suggest that these sources are free-free
absorbed, and may be related to heavily embedded sources in the early stages
of evolution.

\begin{figure} 
\epsscale{1.} 
\plotone{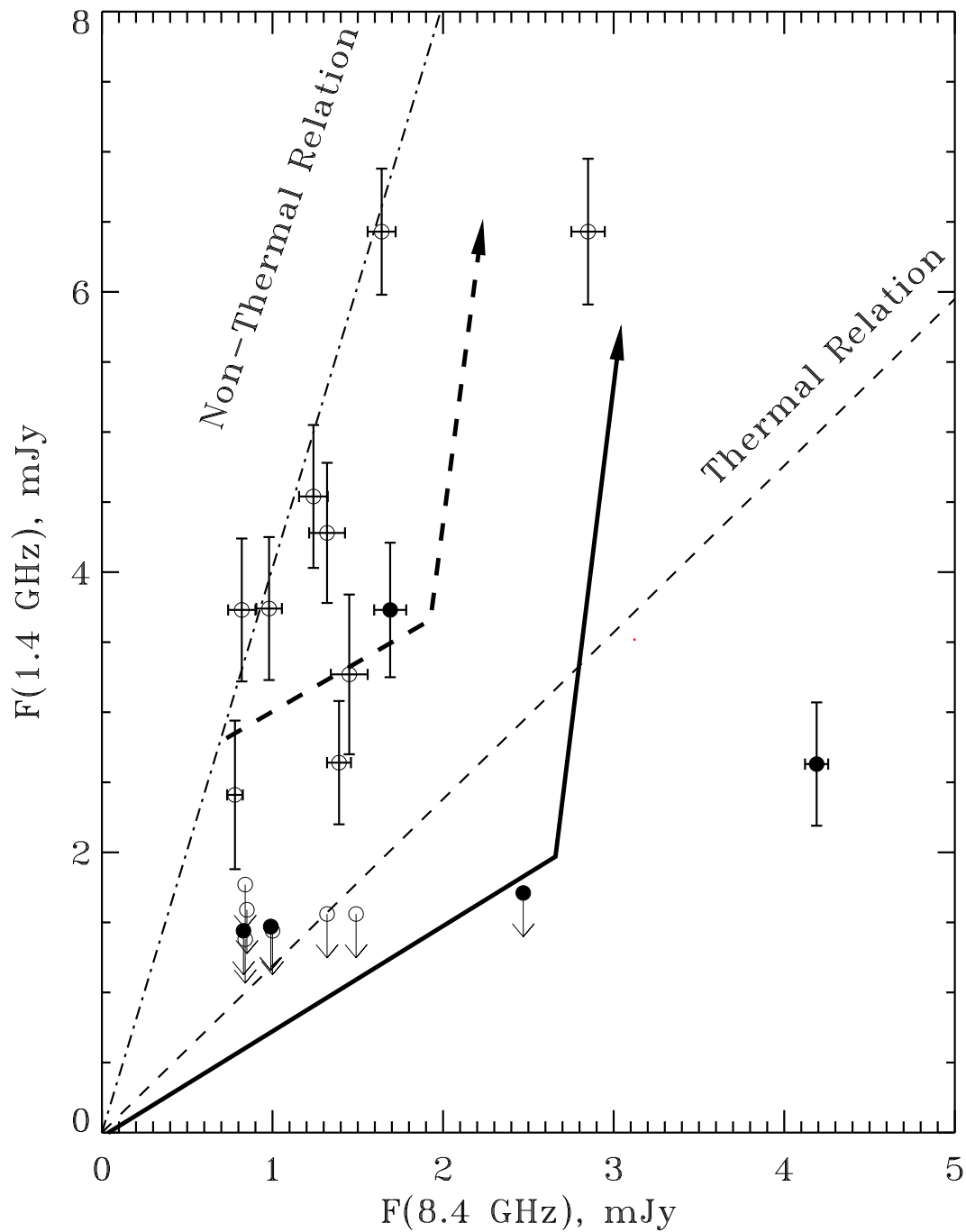} 
\caption{\label{Fig:ThNonTh} 
Distribution of F(1.4~GHz) as a function of F(8.4~GHz) for the galaxies
in Table~\ref{Tab:RadioData}. 
The dashed thin line shows the expected
position of  galaxies dominated by thermal emission, while the
dot-dashed line shows where galaxies dominated by non-thermal
emission are located. Superimposed on this plot are solid-thick 
and dashed-thick lines
indicating qualitative evolutionary tracks described in the text.
} 
\end{figure}

In Figure~\ref{Fig:ThNonTh} we present the distribution of our
galaxies in the F(1.4~GHz) {\it vs.} F(8.4~GHz) diagram.
This Figure also presents two lines indicating the location of
thermal ($\alpha=-0.1$) and  synchrotron emission
($\alpha=-0.8$). In the case of more evolved
star forming galaxies, where we find a mixture of these two
components, the 1.4~GHz emission is dominated by
synchrotron emission, while at 8.4~GHz both synchrotron
and thermal emission have similar contributions. Most of the
galaxies in our sample lie in the region between these
two lines, indicating that they are young objects still dominated
by thermal emission. The position of UM\,533 in this diagram (2.6;4.2), as
well as other galaxies for which we only have upper limits
at 1.4~GHz, is consistent with the description given above.

It is obvious that Figure~\ref{Fig:ThNonTh} represents an oversimplification,
as the history of star formation is much more complicated and cannot be
represented  by either continuous or a single burst or even a couple of
them (e.g.~Terlevich et al.~2004). Keeping this in mind 
we overplot two qualitative evolutionary tracks to the points in
Figure~\ref{Fig:ThNonTh}. In the first case 
(solid thick line) we assumed 
that the galaxies start with only thermal emission from the OB
associations, which are heavily absorbed at both 1.4~GHz and 8.4~GHz
and can only be detected at the highest frequencies. In this phase the
dense giant HII regions are characterized by optically thick 
free-free emission, which is observed as an inverted spectrum (thermal) radio
source and it is commonly named  ultradense HII region 
(Kobulnicky \& Johnson 1999, Johnson et al. 2003). 

Due to the effect of stellar winds, with typical velocities of 
10$^3$ km s$^{-1}$ and mass losses of about 
3 $\times 10^{-5}$\Msolar\ year$^{-1}$ (e.g. Tenorio-Tagle et al. 1990), 
the absorbing material starts to blow away, the optical depth is reduced 
and  the galaxy moves in this diagram towards the direction of  UM\,533 
(2.6;4.2). 
Once all the absorbing material is blown away, the galaxy moves up in
this diagram, until it reaches the line where the radio emission
is dominated by thermal radiation. 
As the galaxy continues to evolve, SNe
explode, producing copious amounts of synchrotron radiation and moving
the galaxies to the top right portion of this diagram where the 
relation between radio and FIR is obtained (Yun, Reddy \& Condon 2001). 
In this scenario we expect young galaxies to have $\alpha>-0.1$,
with the younger objects having the larger slopes due to free-free
absorption, evolving towards $\alpha=-0.8$. 

In the second scenario we start with some synchrotron emission due to a
previous generation of supernovae. If a new burst occurs a new generation of
stars is produced, increasing the amount of ionizing photons, and moving the 
galaxy to the right towards the thermal relation (see dashed thick line in 
Figure~\ref{Fig:ThNonTh}). When the next generation of  supernova explodes
the galaxy starts to move up and right to the position of the non-thermal
relation (dot-dashed line). 
In the next section
we explore further these two scenarios  by including the analysis of
optical data from HIIG91.

\section{Optical Properties of the Sample}
\label{Sec:Opt}

For the observed galaxies we collected optical spectral information from
HIIG91. We corrected by extinction the fluxes of the \Ha\, and \Hb\,  
lines using the Milky Way  extinction curve (Seaton 1979).
The ratios between the observed \OIII\, and \Hb\, and the EW(\Hb) are also
included in the analysis (Table~\ref{Tab:OptData}). 

\begin{figure} 
\epsscale{1.} 
\plotone{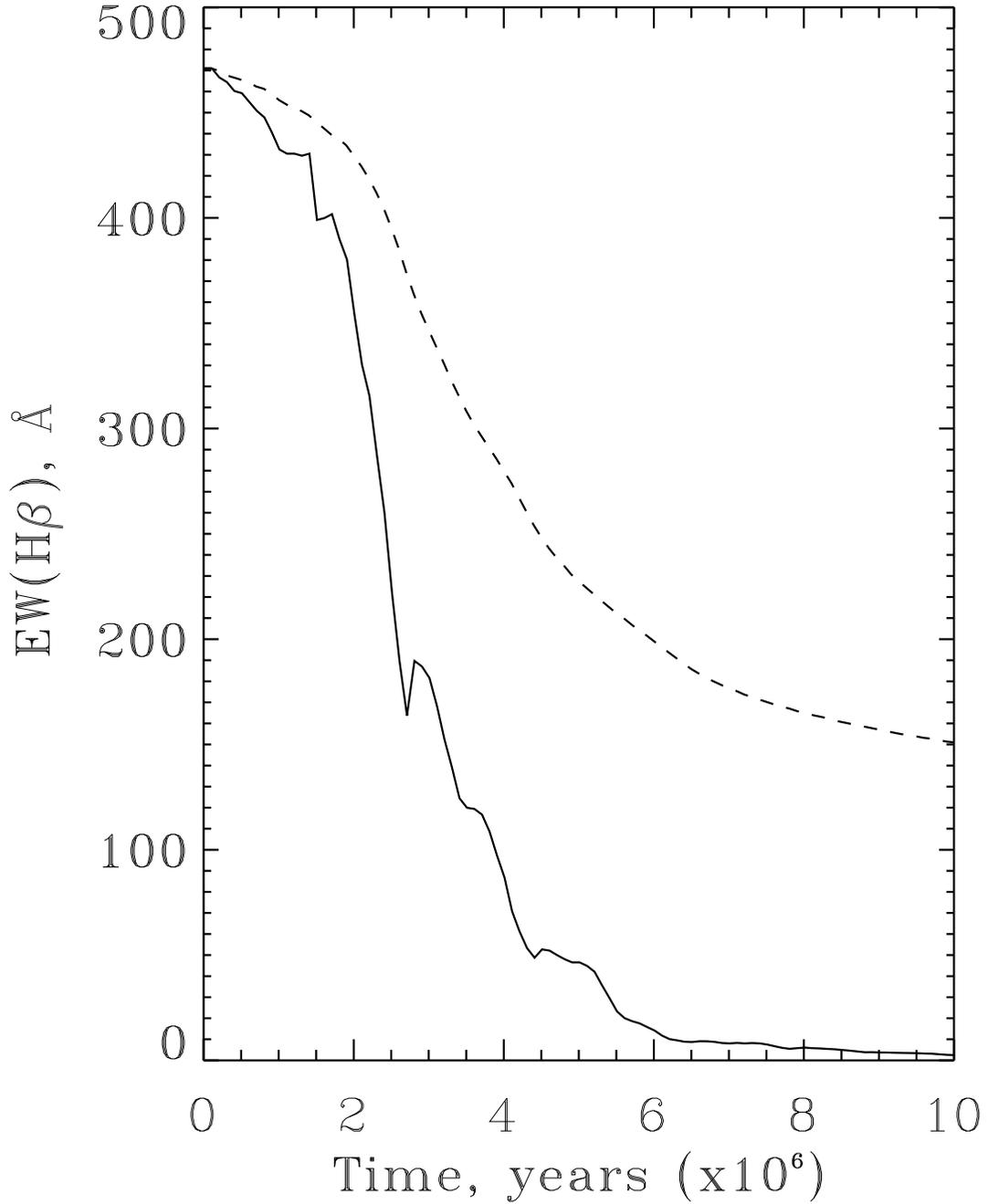} 
\caption{\label{Fig:EWevol} Evolution of the EW(\Hb) for the case of 
an instantaneous burst, solid line, and continuous star formation, 
dashed line (see text).
} 
\end{figure}

The EW(\Hb) has long been used to estimate the age of a stellar burst 
(e.g. Dottori 1981, Luridiana \& Peimbert 2001).
However,  the EW(\Hb) indicator which consists of the relation 
between the continuum flux which depends on the whole 
star formation history of the galaxy and the strength of the emission
line which depends on the recent ($\sim 5\times 10^6$ years) star formation 
activity can  be significantly lower than the age of the current burst.
This problem has been noticed in the analysis of the most extreme  
HII galaxies by Terlevich et al. (2004).
To illustrate this problem, we plot in Figure~\ref{Fig:EWevol} 
the evolution of the EW(\Hb) for the case of an 
instantaneous burst and for the case of continuous star formation. 
Both models were calculated using the SB99 code~\citep{1999Leitherer}
for the case of a Salpeter initial mass function and masses between
0.1 and 100\Msolar. In both modes the EW(\Hb) decreases with time 
but due to the formation of new  massive stars 
in the continuous mode, the EW(\Hb) remains larger as time increases.
These star forming histories  mark the two extrema and the age 
of a  galaxy with a given   EW(\Hb) must lie between the limits
shown in figure~\ref{Fig:EWevol}.
In  the case of the existence of an old population, the observed  EW(\Hb) 
is reduced even more -- in comparison with the continuous case -- 
due to the integrated light of the galaxy and the absence of massive
stars responsible for the recombination lines.

\begin{figure} 
\epsscale{1.} 
\plotone{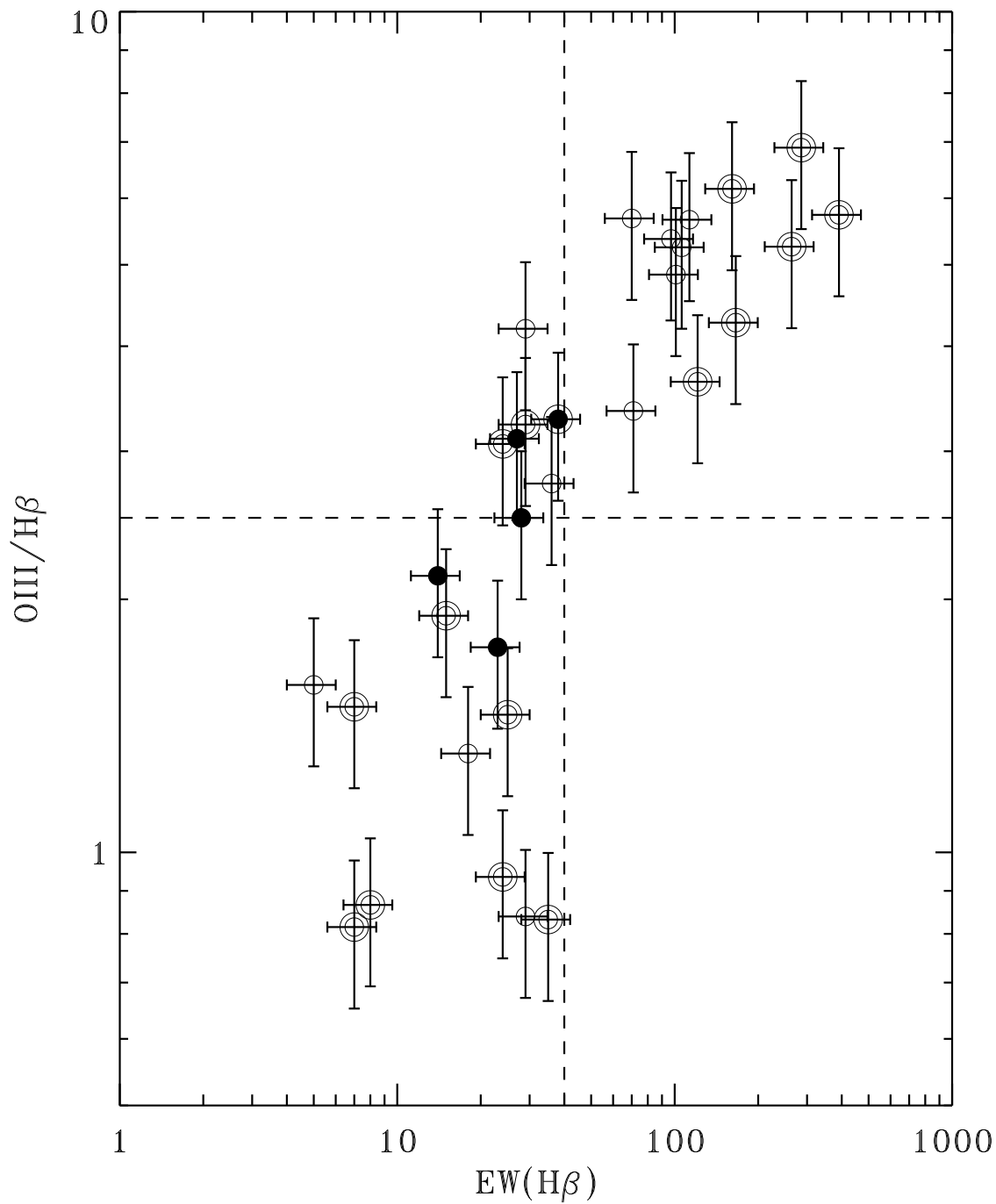} 
\caption{\label{Fig:OpticCorr} Correlation between the EW(\Hb) and the ratio
  OIII/\Hb.  Galaxies with $\alpha^{1.4}_5 <-0.5$ (solid circles) and  
with $\alpha^{1.4}_5 \ge -0.5$ (open
symbols) are represented. Double circles indicate 
those galaxies for which the 20 cm flux is just an upper limit. The vertical
dashed line is at EW(\Hb)=40\AA\ and separates two regions of the plot with
different scatter; the horizontal dashed line separates regions of low from 
high excitation (see description in the text). 
} 
\end{figure}

For our sample we found that there is a good correlation
between the  EW(\Hb) and the  strength of the \OIII\, line  
which is stronger in the early phases of the starburst
when the EW(\Hb) is higher. In fact figure~\ref{Fig:OpticCorr} 
shows that the scatter is higher for those galaxies
with  EW(\Hb) lower than about 40\AA. This figure supports the idea 
that we are mostly dealing with single young bursts on the upper right 
corner of the plot, while the bottom left represents a superposition of 
bursts on an older underlying system. 
\clearpage
Based on the reddening corrected \Ha\,  fluxes
of the galaxies in the HIIG91 catalog, we estimate their Star Formation 
Rates (SFR)
using the relation given by Rosa-Gonz\'alez, Terlevich \&  Terlevich (2002), 
\begin{equation}\label{eq:SFR_Ha}
\rm SFR [M_\odot year^{-1}]  = 1.1\times 10^{-41}\times L(H\alpha) 
\end{equation}
where the L(\Ha) is given in erg s$^{-1}$ and must have been corrected by 
extinction.
The relation given by Equation~\ref{eq:SFR_Ha} was the result of comparing 
different indicators of SFR in a well 
defined sample of starburst galaxies and using as reference the SFR 
given by the FIR. The conversion between \Ha\, luminosities and SFR 
is in any case just 15\% higher than the conversion given by other authors
(e.g. Kennicutt, Tamblyn \& Condon 1994; Madau, Pozzetti  \& Dickinson 1998).
The extinction corrected values of the \Ha\, and 
\Hb\, fluxes  together with the corresponding SFRs are given in Table~\ref{Tab:OptData}.


We make use of the SB99 code to estimate the mass of the current burst. 
In the case of an instantaneous burst, a stellar cluster of 10$^6$ \Msolar\, produces
a luminosity in \Ha\, of about 3.2$\times 10^{40}$ erg s$^{-1}$.
The obtained luminosities and derived masses are  given in Table~\ref{Tab:OptData}.


\section{Combining  Optical and Radio Measurements}
\label{Sec:OpticalRadio}

In this section we combine the radio and optical data described in 
previous sections in order to test the evolutionary models outlined in
section~\ref{Sec:Radio}.

After an episode of star formation, 
the first SN explosion and consequently the first contribution to the
synchrotron  emission appears at about 3.5$\times 10^6$ years,  
which corresponds to an EW(\Hb) of about 120\,\AA\, for the case of an
instantaneous burst or  310\,\AA\, in the case of continuous 
star formation (Figure~\ref{Fig:EWevol}). 
Therefore objects with high EW(\Hb) and flat slopes are
candidates to be galaxies dominated by very recent starburst
episodes.

\begin{figure} 
\epsscale{1.} 
\plotone{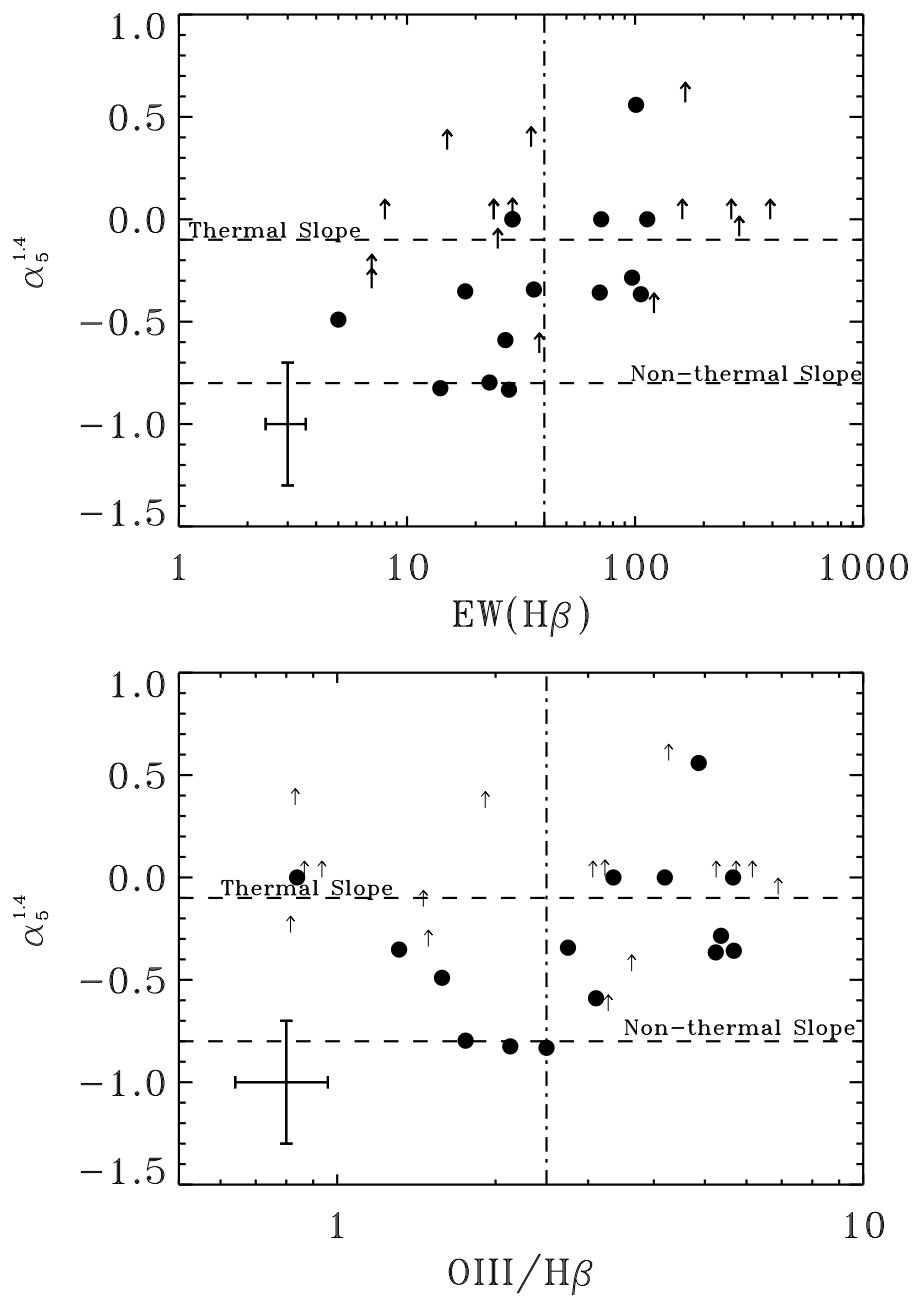} 
\caption{\label{Fig:Slope} Variation of the spectral index ($\alpha^{1.4}_5$) 
with respect to the EW(\Hb) (upper panel) and to the ratio OIII/\Hb\, (lower 
panel). 
In both panels the canonical values for the thermal ($\alpha=-0.1$) 
and non thermal slopes ($\alpha=-0.8$) are represented by dashed lines. 
Dot-dashed lines in both panels reproduce the boundaries indicated by the 
dashed lines in Figure~\ref{Fig:OpticCorr}.
} 
\end{figure}

Figure~\ref{Fig:Slope} shows the relation between the EW(\Hb) and the 
 $\alpha^{1.4}_5$ where objects with EW(\Hb) $> 100$\AA\, have 
slopes higher than -0.5.
A similar relation is found for the case of $\alpha^{1.4}_5$ and
the ratio \OIII/\Hb.
In fact all the galaxies with EW(\Hb) $>$ 40\AA\, and \OIII/\Hb $\gtrsim$ 4 
are flatter than $\alpha^{1.4}_5 > -0.5$ (figure~\ref{Fig:OpticCorr}).
Although Figure~\ref{Fig:Slope} is a scattered plot, there is a tendency
marked by the fact that the points distribution is different in both sides
of the vertical line corresponding to the separation between
``young-simple" and ``old-complex'' systems (Figure~\ref{Fig:OpticCorr}).

For the galaxies with lower EW(\Hb) and detected at 20~cm there is a 
weak trend where galaxies with lower EW(\Hb) have lower values for the 
spectral index $\alpha^{1.4}_5$. Deeper VLA observations of
galaxies  not detected at 20~cm but with EW(\Hb) $\gtrsim$ 5 
(age of about 8$\times 10^6$years for an instantaneous burst)
will be obtained in the near future.
Massive stars within a young burst ionize the surrounding media 
producing free electrons and the presence of recombination lines.
The intensity of the \Hb\, emission line is related to the free-free radiation
observed at radio wavelengths  by (Condon 1991),

\begin{equation}\label{eq:HbRadio}
\rm H\beta\, (erg\, s^{-1}cm^{-2})=  2.8 \times 10^{-11}  (Te/10^4)^{-0.52}  \nu^{0.1}  F_\nu
\end{equation}
where Te is the electron temperature of the plasma,  $\nu$ is the
frequency of the observation (expressed in GHz) and F$_\nu$ is the flux in 
mJy at the given frequency.

\begin{figure} 
\epsscale{.8} 
\plotone{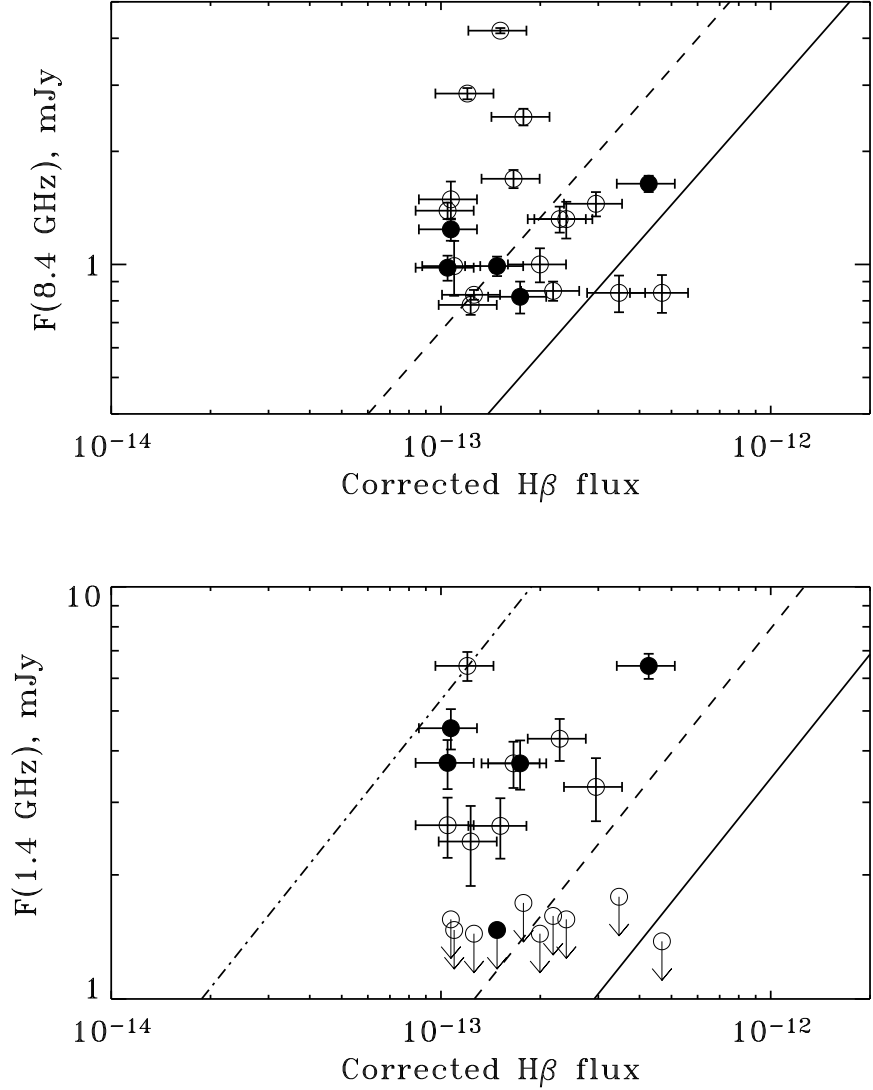} 
\caption{\label{Fig:OptRadio} 
Relation between the extinction corrected \Hb\, flux, the flux at 8.4 GHz 
(top panel)
and the flux at 1.4 GHz (bottom panel).  In both panels the solid line is the 
relation between the strength of the \Hb\, line and the corresponding radio flux
for the case of thermal emission coming from a plasma 
with a temperature of $10^4$K. The dashed lines correspond 
to a temperature of 5$\times 10^4$K.
In the bottom panel we draw (dot-dashed-line) the relation given by 
Yun et al. (2001)  which includes synchrotron emission.
In both panels, the solid symbols are galaxies with $\alpha^{1.4}_5 > -0.5$.
Non detections are represented by the arrows (upper limits).
} 
\end{figure}

For a normal galaxy (normal in the sense that it follows Yun et al.~2001)
dominated by synchrotron radiation, the thermal emission is about 10~percent
of the synchrotron radiation at 1.4GHz. However, due to the different 
behavior and slopes of thermal and non-thermal emission, the thermal emission
at 8.4GHz (3.5cm)
could be about 50~percent of the synchrotron one or even more for 
those galaxies that are synchrotron deficient (Condon 1992;
Schmitt et al. 2006). Figure~\ref{Fig:OptRadio} (top panel) 
shows the extinction corrected \Hb\ flux against the flux at 8.4GHz.
Most of the galaxies have fluxes which are consistent with thermal emission 
from plasmas with temperatures between 10$^4$K (solid line) and  5$ \times
10^4$K (dashed line). 
However some galaxies have an excess of radio emission due to the presence of 
synchrotron emission.

At 20 cm (1.4GHz: bottom panel in Figure~\ref{Fig:OptRadio}) the flux  is
clearly contaminated by the presence of synchrotron emission. 
As in the top panel the solid and dashed lines represent 
the thermal emission based on Equation~\ref{eq:HbRadio}.
All the galaxies detected at 20 cm have fluxes above the thermal limits, 
however the galaxies not detected at 20 cm (the upper limits) 
are close to the thermal region. 

Due to the presence of massive stars in normal and starburst galaxies, there 
is a strong relation between the observed FIR and the flux at 1.4 GHz.
This correlation, which covers several orders of magnitude in
luminosities, is explained by 
the presence of massive stars which heat the dust producing the observed
FIR radiation and, as a consequence of the short life time of these stars, 
they explode as SN producing the synchrotron emission
observed at 1.4 GHz (Yun et al. 2001).
Combining this strong correlation with the relation between 
the FIR luminosities and the current SFR from Kennicutt (1998), 
Yun and collaborators proposed a robust relation between 
the luminosities at 1.4~GHz and the SFR,  

\begin{equation}
\rm SFR [M_\odot year^{-1}]  =  5.9 \times 10^{-29} \times L_{1.4 GHz} 
\end{equation}
where the luminosities are expressed in units of  erg\, s$^{-1}$ Hz$^{-1}$.
Schmitt et al. (2006) found a similar relation using a slightly different
approach. Notice that all SFR estimates have uncertainties,
because each estimator traces different populations and depend on
several assumptions, e.g.~ the initial mass function,
or on the light attenuation which is a strong function of wavelength.

Comparing the SFR estimated by the 1.4~GHz luminosities with those 
given by the recombination lines (Table~\ref{Tab:OptData})
we find that the optical SFR is on average 4 times higher than 
the SFR given by the radio luminosities.
In the bottom panel of Figure~\ref{Fig:OptRadio} we plot a dot-dashed line 
where the optical SFR is equal to the radio SFR based on the 
Yun et al. (2001) relation. Only three galaxies are close to the 
dot-dashed  line showing that most of the observed starbursts still are 
in the early stages of evolution, when not too many
stars have exploded as SN, and as a result
the observed radio SED is dominated by thermal emission.
The most extreme cases are those galaxies with non detection at 20 cm. 
The cases with clear signature of free-free absorption 
and those dominated exclusively  by thermal emission 
are similar to the regions detected by Cannon \& Skillman (2004) within 
NGC\,625, and by Johnson \& Kobulnicky (2003) in He2-10. Observations
with higher angular resolution are necessary to understand the nature of
these sources.

\begin{figure} 
\epsscale{1.0} 
\plotone{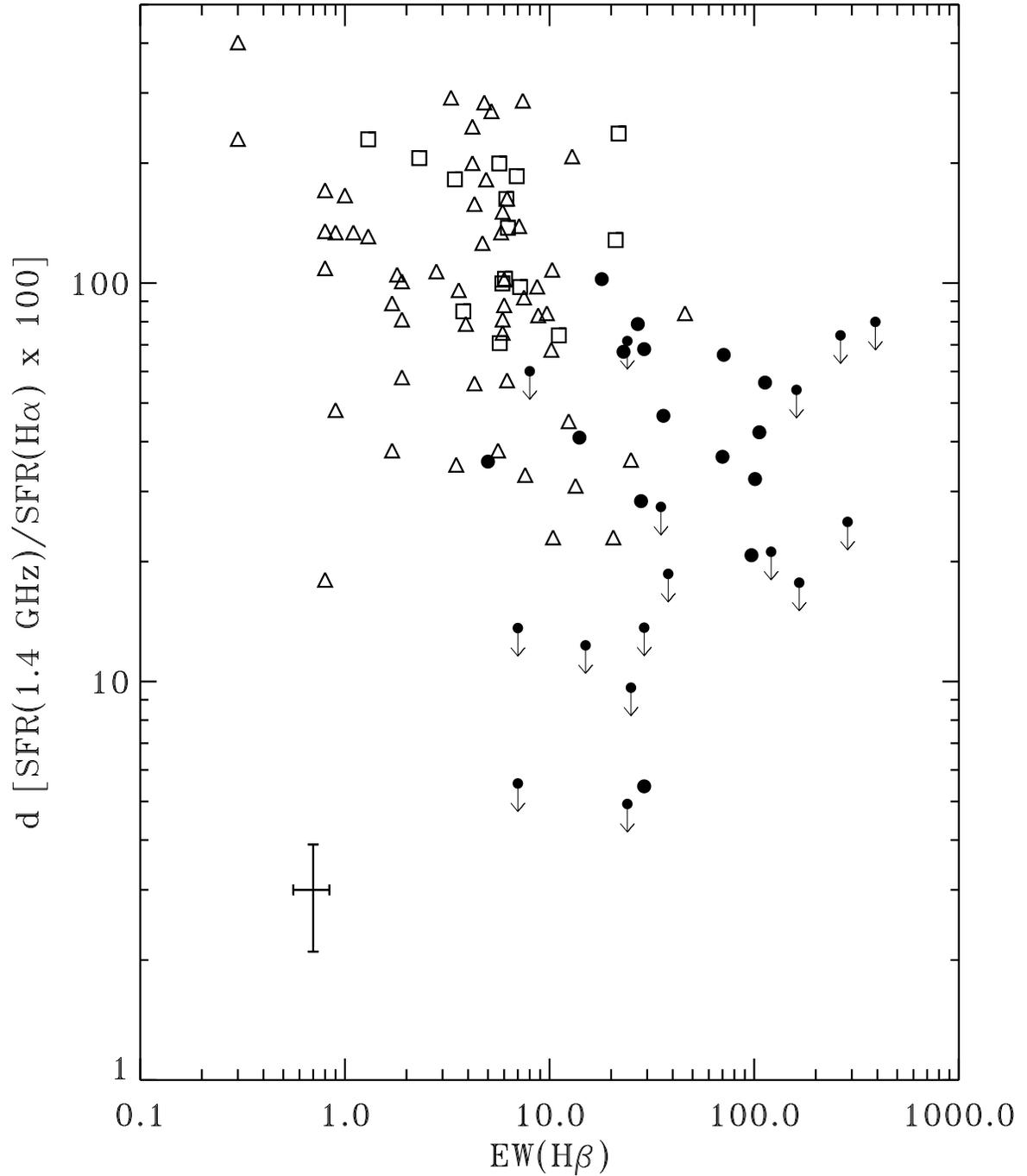} 
\caption{\label{SynchrPercent} \Hb\, equivalent width against the 
$d$ parameter defined as the ratio between the observed
and the expected  1.4 GHz flux (see text for further details). 
Circles are the sample HII galaxies and 
open symbols represent the control sample 
from Jansen et al. (2000) (triangles) and  from Ho et al.~(1997) (squares).
} 
\end{figure}

The deficit of 1.4GHz emission showed by Figure~\ref{Fig:OptRadio} can be
quantified in terms of the ratio between the observed 1.4 GHz flux and the
expected flux based on the SFR calculated using the optical emission lines
and equation (3). For simplicity we denoted this ratio as the $d$-parameter.
Figure~\ref{SynchrPercent} shows the EW(\Hb) against the $d$-parameter
expressed in percentage. Notice that most of our galaxies lie in a region
where the observed flux is less than 50\% of the expected value, showing
that indeed our sample has a deficit of synchrotron radiation.
Five galaxies (Tol\,116-325, MRK\,1315, Tol\,1303-281, Tol\,1304-386 and
Tol\,1358-328) are extreme cases with
the highest EW(\Hb) and just upper limits in the observed 1.4~GHz fluxes.
Notice that, based on the EW(\Hb), these galaxies have ages of less
than 10$^7$ years even in the case of  continuous star formation
(see Figure~\ref{Fig:EWevol}).

In order to investigate how our galaxies compare to normal, quiescent
star forming galaxies in the $d-$parameter~$vs.$EW(H$\beta$) diagram,
we obtained radio 1.4~GHz fluxes (from NVSS) for a sample of such galaxies
from Ho, Filippenko, \& Sargent (1997), and Jansen et al. (2000).
For the Ho et al. (1997) sample we obtained integrated H$\alpha$ fluxes
and EW(H$\beta$) from Moustakas \& Kennicutt (2006) (EW(H$\beta$) was
calculated using the emission line and broad band fluxes).
For the Jansen et al. (2000) sample these values were obtained from Kewley
et al. (2002). The sample, their EW(H$\beta$)'s, SFR(H$\alpha$) corrected
for extinction, SFR(1.4~GHz), F(1.4~GHz) and $d$-parameters are presented
in Table~\ref{Tab:Jansen}. We can see in Figure~\ref{SynchrPercent} that
the normal galaxies have small EW(H$\beta$) and large $d$ values, occupying
a different region relative to the galaxies in our sample, again consistent
with the interpretation that our galaxies are young and have a
deficit of synchrotron radiation. A particularly
interesting result from this Figure is the fact that a significant portion
of the normal galaxies, those with the smaller EW(H$\beta$)'s, have
d$>$100\%. Although this result seems contradictory, since it implies
a higher radio SFR than what is observed in H$\alpha$, it can be explained
in view of the lifetime of these two indicators. For example, in the case of
a single burst, the H$\alpha$ emission lasts only 10~Myr, until all the
ionizing photons die. The radio emission on the other hand will last much
longer than that, because of the longer synchrotron emission life time and
the fact that the stars that explode as SN last for $\sim$40~Myr.

\section{The Far Infrared and the $q$ parameter}
\label{Sec:FIR}

A fraction of the optical-uv light emitted by young and massive stars is 
absorbed by dust grains and re-emitted in the far infrared regime 
making this wavelength range an excellent tracer of the recent star forming 
activity.
 
The $q$ parameter is defined as the ratio between the FIR and radio fluxes 
(Helou, Soifer and Rowan-Robinson 1985), 
\begin{equation}\label{eq:qpar}
q = log \left( \frac{FIR / 3.75\times10^{12}{\rm Hz}}  {S_{1.4 GHz}}   \right)
\end{equation}
to quantify the strong correlation between the radio continuum and the 
FIR observed in a sample of disk-  and starburst- galaxies.

In order to estimate the $q$ parameter of our sample of galaxies we 
obtained from the IRAS catalog (Moshir et al. 1990) 
the fluxes at 60\,$\mu$m  ($S_{60}$) and 100\,$\mu$m ($S_{100}$).
The FIR flux is calculated using the expression from Young et al. (1989), 

\begin{equation}
\label{eq:FIR}
FIR\, ({\rm \, W\, m}^{-2})=1.26 \times 10^{-14} ( 2.58 \times S_{60} + S_{100} )
\end{equation}
where the fluxes are given in Jy (Table~\ref{Tab:Tracers}). 
In those cases where the 100\,$\mu$m flux is just an upper limit we
apply a bolometric correction based on multigrain dust models for the FIR
emission presented by Rowan-Robinson \& Efstathiou (1993). Following this
work, the FIR luminosity is about 1.7 times the luminosity at 60 $\mu$m
(see also Rowan-Robinson et al. 1997, Chapman et al. 2000 for the application 
of this bolometric correction to a different set of starforming galaxies).

\begin{figure} 
\epsscale{.8} 
\plotone{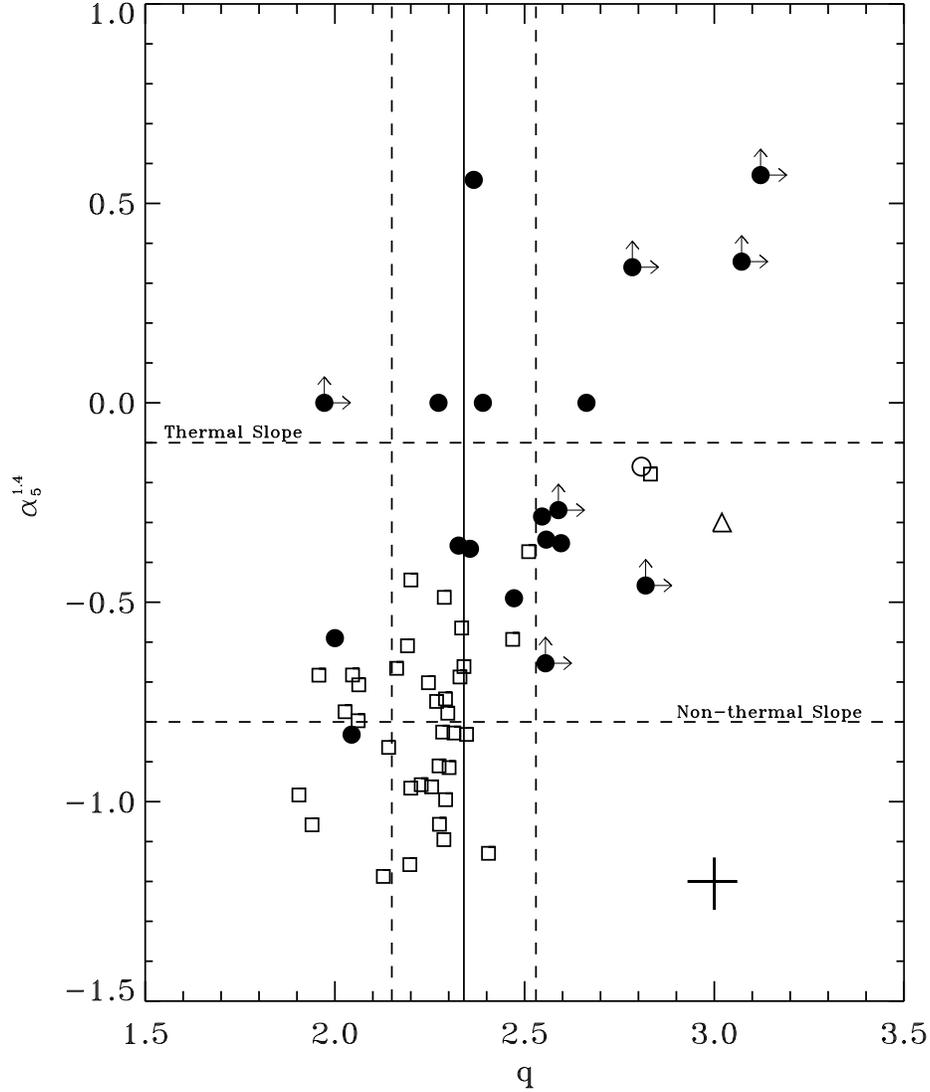} 
\caption{\label{qPar} 
Estimated $q$ parameter for the galaxy sample plotted against the radio spectral index
between 1.4 and 5 GHz (solid circles).  The solid line shows the value of
$q$ found by Roussel et al. (2003).  Two vertical dashed lines
show the one sigma error of this determination. 
Two  well studied {\it nascent}  galaxies are plotted, 
NGC~5253 (open circle, Turner et al. 1998) and NGC~4418 
(triangle, Roussel et al. 2003).
The squares correspond to a sample of normal galaxies extracted 
from Ho et al. (1997).
The horizontal dashed lines show the theoretical values of the slope for the 
cases of thermal and non-thermal emission.    
The cross at the bottom right corner represents a
typical error bar of mean uncertainties of about 
15\% for both IRAS and radio data.} 
\end{figure}

Figure~\ref{qPar} indicates a trend between the $q$ parameter 
and the radio spectral index $\alpha^{1.4}_5$. This figure shows that
galaxies with a $q\lesssim$2.35 -- the canonical value of the 
$q$ parameter (Helou et al.  1985, Roussel et al. 2003) -- are located 
closer to the synchrotron slope confirming the relation 
observed in large samples of normal and starburst galaxies (e.g.~Yun 
et al.~2001).
However galaxies with high $q$ parameter are in general 
close to- or above the line defined by the thermal radio emission slope. 
In some extreme cases the $q$ parameter is higher than 2.75. Such extreme 
galaxies  have been reported in the literature  
(e.g. NGC~1377, NGC~4491, NGC~4418, Roussel et al. 2003).
A high value of  the $q$ parameter  and a positive radio spectral index
point to the presence of free-free absorption as has been 
observed in very young systems 
(Tarchi et al. 2000, Vacca, Johnson \& Conti 2002, 
Johnson \& Kobulnicky 2003).  

In order to compare our results with a large sample of galaxies, we analyzed
the distribution of HII galaxies from Ho, Filippenko \& Sargent (1997) in the
$q$ vs. $\alpha^{1.4}_5$ plot. These galaxies span a wide range of 
characteristics,
from sources similar to the ones studied in this paper to quiescent star
forming galaxies. 
Table~\ref{Tab:Ho-Sample} presents the galaxies from Ho et al. (1997) for 
which we could find integrated 1.4~GHz, 4.9~GHz and IRAS fluxes in the
literature. The 1.4~GHz fluxes were
obtained from NVSS, requiring the addition of multiple components in the 
case of the most extended sources. For the 4.9~GHz we used data from 
single dish observations, mostly from Gregory \& Condon (1991), which had a 
resolution comparable to
that of NVSS.  Using Equation~\ref{eq:qpar}
we estimate the $q$ parameter and plot these results
with our galaxies in Figure 9. As expected, based on previous studies of normal
galaxies (e.g. Condon 1992), most of the galaxies lie in a region 
characterized by a $q$ parameter around 2.5, and  
synchrotron like  slope ($\alpha^{1.4}_5\la -0.5$).  
Only NGC~4214 has characteristics of a young galaxy
and it is located close to NGC~5253.
Notice also that, when the Ho et al. (1997) galaxies are included in this
plot, a much clearer trend emerges. Galaxies with large $q$ parameters tend
to have large $\alpha^{1.4}_5$ slopes, strengthening the idea that they are
young sources in the early stages of evolution. 
Conversely sources with lower $q$ values
usually have smaller $\alpha^{1.4}_5$ slopes, indicating that they are at a
much more advanced evolutionary stage.

\section{Discussion and Conclusions}
\label{Sec:Discussion}

The tight correlation between far infrared (FIR)
and radio luminosity confirms that radio emission is also
a good tracer of recent star formation (Yun et al. 2001).
The origin of this correlation is the population of massive stars that
are responsible for the dust heating, the FIR radiation and the supernova 
activity that generates the radio emission.

However, deviations from this correlation are expected both in the early
phases of a starburst -- before the first Type II SNe explode -- or
in the post-starburst phase when the synchrotron emission ages and is also
contaminated by the emission from less massive stars (Bressan et al.
2002).  

Based on the combination of optical, FIR and radio data, 
we have detected a large deviation from the FIR-radio  correlation 
within a sample of normal and starburst galaxies.
Our sample of 31 galaxies is strongly biased to young systems which have 
strong EW(\Hb) and relatively high SFR. Only a few galaxies in
our sample follow the Yun et al. (2001) relation and the
SFR calculated from radio continuum at 1.4 GHz 
is a factor of 4 below the SFR given by optical emission lines. 
We quantified this deficit introducing the $d$-parameter which 
is the ratio between the observed radio continuum flux at 1.4 GHz
and the expected one calculated 
using the SFR provided by the optical emission lines (H$\alpha$)
and the empirical relation given by Yun et al. (2001) which relates
the flux at 1.4GHz with the current SFR determined by the FIR emission.
In most of the cases the $d$-parameter
is well below 50\% with only 2 galaxies having values above 75\%.

The summary of our results is presented in Table~\ref{Tab:Summary}. 
We tabulated the extreme values of the parameters which point to 
the existence of a recent star forming event in our sample of galaxies. 
The indicators with high values of the EW(\Hb) and \OIII /\Hb\ ratio,
have a value of $\alpha^{1.4}_5$ typical of a region dominated by
thermal emission, a low $d$ parameter, and a high $q$ parameter.
Only two galaxies, Mrk\,1318 and Tol\,1358-328 have the five indicators 
corresponding to very young objects 
while 5 galaxies present 4 of the 5 indicators. 
In the galaxies Tol\,0226-390, Tol\,1247-232 and UM\,533 the only 
indicator not pointing to a very young system is the $q$ parameter 
which is in all cases very close 
to the canonical value. For the case of Tol\,1304-386 we have no 
FIR information so we could not estimate the $q$ parameter. Mrk\,711 has a 
relatively low EW(\Hb)
but a high \OIII /\Hb\ ratio, 
pointing to the existence of a new burst on top of an
older stellar population. In this galaxy all the other tracers of a young
population are present. Another 10 galaxies show signs of young systems 
in 3 of the
5 tracers considered (see Table~\ref{Tab:Summary}) and are also extremely
good candidates to look for deeply embedded super stellar clusters where
only thermal radio and mid infrared radiation can be detected.

For the early phase deviation to occur, a highly synchronized 
star formation is required, i.e. the massive star forming phase
should be shorter than $\sim$3-4Myr in order to prevent the appearance of 
the first supernovae
associated to the more massive stars and the consequent synchrotron emission.
This also suggests that there has been no large starforming event in the
previous $10^8$yr.
The results summarized above have important implications not only for 
low mass galaxy evolution but also for the study of
young massive star clusters and for the understanding of the early stages of
their evolution. We showed that by combining optical spectroscopy with
radio continuum observations we can efficiently detect extremely young
systems, before the first Type II SNe explode. 
These systems represent a significant enlargement of the current sample
of nearby starbursts, like Henize\,2-10 (Johnson \& Kobulnicky 2003)
and NGC\,5253 (Beck et al. 1996, Turner et al. 2000, 
Turner \& Beck 2004), known to have these properties.
A detailed higher resolution study of these sources, to be published
in a forthcoming paper, will be extremely valuable
for the understanding of their properties and to study evolutionary 
scenarios. Furthermore, we hope that 
the study of a complete sample will allow the 
segregation between instantaneous and extended starformation scenarios.

\acknowledgments

The National Radio Astronomy Observatory is a facility of the National
Science Foundation, operated under cooperative agreement by Associated
Universities, Inc. This research made use of the NASA/IPAC Extragalactic
Database (NED), which is operated by the Jet Propulsion Laboratory, Caltech,
under contract with NASA. Basic research at the US Naval Research Laboratory
is supported by the Office of Naval Research.

Computer facilities for DRG were kindly provided by 
{\it Dos-Inform\'atica}, 
Tenerife. RT and ET acknowledge support by the Mexican
research council (CONACYT) under grant  40018. HRS thanks the hospitality of 
INAOE for a visit during a Guillermo Haro Advanced Program in Astrophysics 
workshop when this project was conceived. The authors are very grateful to an 
anonymous referee whose comments and suggestions largely improved this 
manuscript.





\clearpage

\begin{deluxetable}{lrrrrrrrrrl}
\tablewidth{0pt}\tabletypesize{\scriptsize}
\tablecaption{\label{Tab:RadioReconst}Sample and Image Reconstruction Parameters}
\tablehead{
\colhead{Name} & \colhead{RA} & \colhead{DEC} & \colhead{V$_{Rad}$} &\multicolumn{2}{c}{1.4~GHz}&\multicolumn{2}{c}{4.9~GHz}
&\multicolumn{2}{c}{8.4~GHz}&\colhead{}\\
\cline{2-3} \cline{5-6} \cline{7-8} \cline{9-10}\\
\colhead{}     & \colhead{}   & \colhead{}    & \colhead{}          &\colhead{b$_{maj}\times b_{min}$} & \colhead{PA}
&\colhead{b$_{maj}\times b_{min}$} & \colhead{PA}&
\colhead{b$_{maj}\times b_{min}$} & \colhead{PA}& \colhead{Program\tablenotemark{a}}\\
\colhead{} & \multicolumn{2}{c}{(J2000.0)} & \colhead{(km s$^{-1}$)} & \colhead{(\arcsec )} & \colhead{($^{\circ}$)}
& \colhead{(\arcsec )} & \colhead{($^{\circ}$)}& \colhead{(\arcsec )} & \colhead{($^{\circ}$)} & \colhead{}}
\startdata
UM\,304         &01 06 54.1  &+01 56 44   & 4708 &  54.1$\times$47.5 &  40.8 & 18.3$\times$14.6 &  54.4 & 10.3$\times$8.5  &  61.2 & AG674\\
Tol\,0127-397   &01 29 15.9  &-39 30 38   & 4797 &  50.4$\times$45.4 &  30.7 & 43.0$\times$12.6 &   6.5 &  2.8$\times$0.6  &  -8.1 & AG674,AS734\\
Tol\,0145-391   &01 47 10.3  &-38 52 53   &16250 &  54.4$\times$47.1 & -24.2 & 40.8$\times$12.7 &   6.7 & 24.1$\times$7.4  &  11.4 & AG674\\
Tol\,0226-390   &02 28 12.3  &-38 49 21   &14330 &  52.9$\times$46.9 &  46.9 & 42.8$\times$12.9 &  10.9 & 22.9$\times$7.3  &   6.4 & AG674\\
MRK\,605        &03 15 37.3  &-03 28 11   & 8410 &  51.1$\times$46.7 & -21.9 & 18.5$\times$14.8 &  50.6 & 12.0$\times$8.6  &  48.3 & AG674\\
Tol\,0614-375   &06 16 13.8  &-37 36 37   & 9893 &  65.1$\times$46.2 &  31.1 & 42.1$\times$16.0 & -20.3 & 26.9$\times$7.7  & -22.8 & AG674\\
Tol\,0619-392   &06 21 03.6  &-39 17 07   & 1588 &  57.3$\times$48.2 & -44.5 & 46.1$\times$14.4 & -14.9 & 25.0$\times$8.3  & -10.7 & AG674\\
MRK\,711        &09 55 11.4  &+13 25 46   & 5756 &  57.0$\times$49.4 &  44.1 & 32.2$\times$14.3 & -70.8 & 20.6$\times$7.7  & -68.5 & AG674\\
Tol\,0957-278   &09 59 21.2  &-28 08 00   &  710 &  74.1$\times$53.0 & -36.6 & 10.3$\times$2.2  & -50.3 & 40.9$\times$10.6 & -46.8 & AS412,AG674\\
MRK\,36         &11 04 58.5  &+29 08 22   &  646 &  53.7$\times$50.0 &  76.7 &  4.4$\times$4.2  &  13.4 &  8.4$\times$8.2  &  48.7 & AK331,AS734\\
Tol\,1116-325   &11 18 28.7  &-32 53 10   &  600 &  92.4$\times$85.7 &  14.4 & 33.8$\times$16.9 &   6.8 & 17.7$\times$9.4  &  -0.8 & AG674\\
UM\,421         &11 20 10.7  &-00 16 32   &12039 &  58.0$\times$47.8 &  28.2 & 46.6$\times$14.4 & -61.0 & 30.0$\times$8.1  & -60.1 & AG674\\
UM\,444         &11 40 13.2  &-00 24 42   & 6598 &  55.3$\times$50.0 & -25.9 & 47.4$\times$14.3 & -58.6 & 24.6$\times$8.6  & -61.7 & AG674\\
UM\,468         &11 55 59.2  &-01 00 01   &10931 &  57.8$\times$47.7 & -13.3 & 40.6$\times$14.6 & -61.4 & 26.1$\times$8.7  & -61.1 & AG674\\
UM\,482         &12 12 03.3  &-00 36 22   &10571 &  50.0$\times$47.3 &  24.1 & 41.4$\times$14.9 & -62.0 & 21.3$\times$8.7  & -65.0 & AG674\\
MRK\,1315       &12 15 18.7  &+20 25 03   &  847 &  75.5$\times$49.3 & -73.7 &  1.4$\times$1.2  & -60.9 & 17.0$\times$8.1  & -75.4 & AS286,AG674 \\
MRK\,1318       &12 19 09.9  &+03 51 21   & 1526 &  53.5$\times$46.8 &  -2.2 &  4.8$\times$4.3  &   5.7 &  3.0$\times$2.5  &  43.9 & AR541\\
UM\,488         &12 18 26.7  &-00 07 50   &15115 &  52.7$\times$48.9 &  58.5 & 30.7$\times$15.8 & -69.1 & 16.6$\times$9.1  & -71.8 & AG674\\
MRK\,52         &12 25 42.8  &+00 34 22   & 2140 &  52.6$\times$50.1 &  -1.4 &  5.4$\times$4.5  & -14.0 & 16.8$\times$9.5  & -72.5 & AW137,AG674 \\
Tol\,1223-388   &12 26 09.8  &-39 07 26   & 3661 &  49.1$\times$45.3 &  24.1 & 46.4$\times$15.0 & -11.4 & 26.4$\times$8.8  & -11.4 & AG674\\
Tol\,1247-232   &12 50 18.9  &-23 33 58   &14390 &  51.6$\times$47.2 &  40.1 & 25.0$\times$18.9 &  -6.1 & 13.9$\times$11.0 &   9.9 & AG674\\
UM\,530         &12 58 08.4  &+01 51 44   &20019 &  60.7$\times$59.7 &  -9.7 & 28.3$\times$16.2 & -70.2 & 17.7$\times$9.0  & -71.3 & AG674\\
UM\,533         &12 59 58.1  &+02 02 59   &  886 &  84.4$\times$49.9 &  16.4 & 27.7$\times$16.1 & -73.2 & 15.0$\times$9.1  & -79.2 & AG674\\
Tol\,1303-281   &13 05 43.7  &-28 25 03   & 1499 &  49.7$\times$48.9 &  -4.1 & 28.6$\times$18.1 &   6.1 & 15.5$\times$9.8  &  -4.5 & AG674\\
Tol\,1304-386   &13 07 21.0  &-38 54 49   & 4197 &  48.1$\times$45.7 &  16.2 & 42.3$\times$13.7 &  -3.1 & 24.0$\times$8.8  &   2.2 & AG674\\
Tol\,1324-276   &13 27 06.6  &-27 57 23   & 2023 &  53.4$\times$46.9 &  23.0 &  2.5$\times$1.1  &  -2.9 &  1.5$\times$0.6  &   4.1 & AK431\\
UM\,603         &13 41 37.7  &-00 25 55   &17988 &  94.3$\times$53.5 &  17.3 & 22.2$\times$15.6 &  57.6 & 18.1$\times$9.4  &  62.9 & AG674\\
Tol\,1358-328   &14 01 21.6  &-33 03 50   & 1216 & 106.8$\times$82.4 &  14.2 & 31.9$\times$16.3 &   2.9 & 17.8$\times$9.8  &  -5.4 & AG674\\
Tol\,2259-398   &23 02 22.5  &-39 33 32   & 8858 &  55.1$\times$47.4 &  34.8 & 44.9$\times$12.3 &   4.9 & 25.0$\times$7.3  &   9.9 & AG674\\
Tol\,2306-400   &23 08 59.6  &-39 45 51   &19487 &  75.0$\times$62.1 & -12.0 & 51.1$\times$12.8 &  17.3 & 25.7$\times$7.4  &  12.6 & AG674\\
MRK\,930        &23 31 58.3  &+28 56 50   & 5485 &  53.7$\times$48.2 & -70.8 &  1.6$\times$1.2  & -43.4 &  9.5$\times$7.6  &  63.8 & AS314,AG674 \\
\enddata
\tablenotetext{a}{AG674 corresponds to our observations. In the case when 
observations were
obtained from 2 different programs, the first one listed corresponds to 
4.9~GHz and the second one to 8.4~GHz.}
\end{deluxetable}

\begin{deluxetable}{lrrrrrl}
\tablewidth{0pt}\tabletypesize{\scriptsize}
\tablecaption{\label{Tab:RadioData}Radio properties of our sample}
\tablehead{
\colhead{Name} & \colhead{F(1.4~GHz)} & \colhead{F(4.9~GHz)} & \colhead{F(8.4~GHz)}
& \colhead{$\alpha^{1.4}_5$}  & \colhead{$\alpha^{5}_8$} & \colhead{Comments}\\
\colhead{}&\colhead{(mJy)}&\colhead{(mJy)}&\colhead{(mJy)}&\colhead{}&\colhead{}&\colhead{}}
\startdata
UM\,304         & 6.43 $\pm$ 0.52 & 4.14 $\pm$ 0.12 & 2.85 $\pm$ 0.10 & -0.35 $\pm$ 0.16 & -0.68 $\pm$ 0.19 & \\
Tol\,0127-397   & $<$1.47         & 0.67 $\pm$ 0.13 & 0.99 $\pm$ 0.06 & $>$-0.65         &\nodata           &a \\
Tol\,0145-391   & $<$1.41         & $<$0.35         & $<$0.27         &\nodata           &\nodata           & \\
Tol\,0226-390   & 3.73 $\pm$ 0.48 & 2.36 $\pm$ 0.13 & 1.69 $\pm$ 0.09 & -0.37 $\pm$ 0.26 & -0.61 $\pm$ 0.33 & \\
MRK\,605        & $<$1.59         & 1.15 $\pm$ 0.07 & 0.85 $\pm$ 0.05 & $>$-0.27         & -0.55 $\pm$ 0.36 & \\
Tol\,0614-375   & $<$1.74         & $<$0.50         & $<$0.25         &\nodata           &\nodata           & \\
Tol\,0619-392   & $<$1.56         & 2.35 $\pm$ 0.23 & 1.32 $\pm$ 0.15 & $>$ 0.34         & -1.05 $\pm$ 0.62 & \\
MRK\,711        & 2.64 $\pm$ 0.44 & 1.72 $\pm$ 0.19 & 1.39 $\pm$ 0.07 & -0.34 $\pm$ 0.37 & -0.39 $\pm$ 0.52 & \\
Tol\,0957-278   & 6.01 $\pm$ 0.51 & 2.00 $\pm$ 0.12 & 2.43 $\pm$ 0.12 &\nodata           &\nodata           &b \\
MRK\,36         & 2.41 $\pm$ 0.53 & 1.54 $\pm$ 0.08 & 0.78 $\pm$ 0.05 & \nodata          & -1.24 $\pm$ 0.33 &c \\
Tol\,1116-325   & $<$1.71         & $<$0.31         & $<$0.18         &\nodata           &\nodata           & \\
UM\,421         & 4.28 $\pm$ 0.50 & 2.32 $\pm$ 0.16 & 1.32 $\pm$ 0.11 & -0.49 $\pm$ 0.25 & -1.03 $\pm$ 0.44 & \\
UM\,444         & $<$1.44         & 1.45 $\pm$ 0.14 & 1.00 $\pm$ 0.10 & $>$ 0.01         & -0.68 $\pm$ 0.60 & \\
UM\,468         & 4.54 $\pm$ 0.51 & 2.17 $\pm$ 0.11 & 1.24 $\pm$ 0.08 & -0.59 $\pm$ 0.23 & -1.02 $\pm$ 0.36 & \\
UM\,482         & $<$1.44         & $<$0.35         & $<$0.25         &\nodata           &\nodata           & \\
MRK\,1315       & $<$1.41         & 0.56 $\pm$ 0.11 & 0.90 $\pm$ 0.08 &\nodata           &\nodata           &b \\
MRK\,1318       & $<$1.44         & 0.83 $\pm$ 0.06 & 0.83 $\pm$ 0.03 & $>$-0.46         &  0.00 $\pm$ 0.34 & \\
UM\,488         & 3.73 $\pm$ 0.51 & 1.33 $\pm$ 0.10 & 0.82 $\pm$ 0.08 & -0.83 $\pm$ 0.28 & -0.88 $\pm$ 0.51 & \\
MRK\,52         & 13.21$\pm$ 0.45 & 4.27 $\pm$ 0.13 & 4.44 $\pm$ 0.09 &\nodata           &\nodata           &b \\
Tol\,1223-388   & $<$1.56         & 2.39 $\pm$ 0.18 & 1.49 $\pm$ 0.17 & $>$ 0.35         & -0.86 $\pm$ 0.58 & \\
Tol\,1247-232   & 3.27 $\pm$ 0.57 & 2.29 $\pm$ 0.13 & 1.45 $\pm$ 0.11 & -0.29 $\pm$ 0.34 & -0.83 $\pm$ 0.39 & \\
UM\,530         & 6.43 $\pm$ 0.45 & 2.27 $\pm$ 0.10 & 1.64 $\pm$ 0.08 & -0.83 $\pm$ 0.15 & -0.59 $\pm$ 0.28 & \\
UM\,533         & 2.63 $\pm$ 0.44 & 5.29 $\pm$ 0.08 & 4.19 $\pm$ 0.07 &  0.56 $\pm$ 0.31 & -0.43 $\pm$ 0.09 & \\
Tol\,1303-281   & $<$1.38         & $<$0.71         & $<$0.35         &\nodata           &\nodata           & \\
Tol\,1304-386   & $<$1.47         & 1.33 $\pm$ 0.22 & 0.99 $\pm$ 0.17 & $>$-0.08         & -0.54 $\pm$ 0.98 & \\
Tol\,1324-276   & 3.90 $\pm$ 0.47 & 0.78 $\pm$ 0.04 & 0.51 $\pm$ 0.03 &\nodata           & -0.78 $\pm$ 0.33 &c \\
UM\,603         & 3.74 $\pm$ 0.51 & 1.38 $\pm$ 0.09 & 0.98 $\pm$ 0.08 & -0.80 $\pm$ 0.28 & -0.62 $\pm$ 0.42 & \\
Tol\,1358-328   & $<$1.71         & 3.40 $\pm$ 0.17 & 2.47 $\pm$ 0.13 & $>$ 0.57         & -0.58 $\pm$ 0.30 & \\
Tol\,2259-398   & $<$1.77         & 1.49 $\pm$ 0.14 & 0.84 $\pm$ 0.09 & $>$-0.14         & -1.05 $\pm$ 0.62 & \\
Tol\,2306-400   & $<$1.38         & 0.92 $\pm$ 0.12 & 0.84 $\pm$ 0.10 & $>$-0.34         & -0.17 $\pm$ 0.74 & \\
MRK\,930        & 11.53$\pm$ 0.48 & 2.72 $\pm$ 0.09 & 3.66 $\pm$ 0.05 &\nodata           &\nodata           &b \\
\enddata
\tablenotetext{.}{Flux upper limits and $\alpha^{1.4}_5$ lower limits are defined as three times
the image r.m.s. level.}
\tablenotetext{a}{Observations at 1.4~GHz and 4.9~GHz were obtained 
with similar resolution, but 8.4~GHz was
observed with too high a resolution to allow a reliable determination 
of $\alpha^{5}_8$.}
\tablenotetext{b}{Observations taken with different resolutions to allow 
the reliable determination of any spectral index.}
\tablenotetext{c}{Observations taken with similar resolution at 4.9~GHz and 
8.4~GHz, but not at 1.4~GHz, so $\alpha^{1.4}_5$ cannot be 
reliably determined.}
\end{deluxetable}

\begin{deluxetable}{lrccccrrcccc}
\tablewidth{0pt}\tabletypesize{\scriptsize}
\tablecaption{\label{Tab:OptData}Optical properties and SFRs of our sample}
\tablehead{
\colhead{Name}      & \colhead{Log F(\Ha)}   & \colhead{Log F(\Hb)} & \colhead{EW(\Hb)}              
& \colhead{A$_{\rm v}$}  & \colhead{[OIII]/\Hb} & \colhead{SFR(H$\alpha$)} & \colhead{SFR(1.4~GHz)} 
& \colhead{Log L(\Ha) } & \colhead{Log Mass } \\
\cline{2-3} \cline{7-8}\\
\colhead{ }    & \multicolumn{2}{c}{(erg s$^{-1}$ cm$^{-2} $)}
& \colhead{(\AA)}& \colhead{ }  & \colhead{ } & \multicolumn{2}{c}{(\Msolar year$^{-1}$)}
& \colhead{(erg s$^{-1}$) } & \colhead{(\Msolar)} \\
}
\startdata
       UM\,304   &  -12.47 &  -12.92 &   18 &  2.33 &  1.31 &   2.1 &   2.1   &  41.28 &    6.78 \\
Tol\,0127-397    &  -12.38 &  -12.83 &   38 &  1.09 &  3.28 &   2.6 &   0.5   &  41.38 &    6.88 \\
Tol\,0145-391    &  -11.34 &  -11.80 &   24 &  0.01 &  0.94 & 312.2 & $<$ 5.1 &  43.45 &    8.95 \\
Tol\,0226-390    &  -12.32 &  -12.78 &  106 &  0.99 &  5.24 &  25.5 &  10.7   &  42.36 &    7.86 \\
      MRK605     &  -12.21 &  -12.66 &    7 &  2.87 &  0.81 &  13.5 &   1.9   &  42.09 &    7.59 \\
Tol\,0614-375    &  -12.33 &  -12.79 &    8 &  2.68 &  0.87 &  12.2 & $<$ 2.5 &  42.05 &    7.55 \\
Tol\,0619-392    &  -12.17 &  -12.62 &   15 &  2.76 &  1.91 &  46.0 &   5.7   &  42.62 &    8.12 \\
      MRK711     &  -12.52 &  -12.98 &   36 &  0.82 &  2.75 &   2.6 &   1.2   &  41.38 &    6.88 \\
Tol\,0957-278    &  -12.33 &  -12.79 &   29 &  1.02 &  4.20 &   0.3 & $<$ 0.2 &  40.41 &    5.91 \\
       MRK36     &  -12.45 &  -12.91 &   70 &  0.96 &  5.68 &   0.1 & $<$ 0.1 &  39.84 &    5.34 \\
Tol\,1116-325    &  -12.43 &  -12.89 &  264 &  1.07 &  5.25 &  0.04 & $<$ 0.1 &  39.51 &    5.01 \\
       UM\,421   &  -12.19 &  -12.64 &    5 &  4.24 &  1.58 &  25.0 &   8.9   &  42.35 &    7.85\\
       UM\,444   &  -12.24 &  -12.70 &   29 &  1.69 &  3.23 &   7.9 &   1.1   &  41.85 &    7.35 \\
       UM\,468   &  -12.52 &  -12.97 &   27 &  1.26 &  3.11 &  10.1 &   8.1   &  41.97 &    7.47 \\
       UM\,482   &  -12.49 &  -12.94 &   24 &  1.49 &  3.06 &  10.2 & $<$ 2.4 &  41.96 &    7.46 \\
     MRK1315     &  -12.55 &  -13.01 &  391 &  0.01 &  5.73 &  0.03 & $<$ 0.1 &  39.40 &    4.90 \\
     MRK1318     &  -12.44 &  -12.90 &  121 &  0.40 &  3.63 &   0.2 & $<$ 0.1 &  40.30 &    5.80 \\
       UM\,488   &  -12.31 &  -12.76 &   14 &  2.33 &  2.13 &  30.9 &  12.6   &  42.45 &    7.95 \\
       MRK52     &  -10.89 &  -11.34 &   29 &  3.06 &  0.84 &  19.9 & $<$ 1.1 &  42.26 &    7.76 \\
Tol\,1223-388    &  -12.52 &  -12.97 &   35 &  1.21 &  0.83 &   0.9 &   0.2   &  40.91 &    6.41 \\
Tol\,1247-232    &  -12.07 &  -12.53 &   97 &  0.74 &  5.37 &  47.4 &   9.8   &  42.63 &    8.13 \\
       UM\,530   &  -11.92 &  -12.37 &   28 &  2.12 &  2.50 & 143.9 &  40.9   &  43.12 &    8.62 \\
       UM\,533   &  -12.36 &  -12.82 &  101 &  1.66 &  4.87 &   0.2 &   0.1   &  40.19 &    5.69 \\
Tol\,1303-281    &  -12.39 &  -12.84 &  161 &  1.15 &  6.16 &   0.2 & $<$ 0.1 &  40.35 &    5.85 \\
Tol\,1304-386    &  -12.50 &  -12.96 &  286 &  0.67 &  6.89 &   1.5 &   0.4   &  41.13 &    6.63 \\
Tol\,1324-276    &  -12.43 &  -12.88 &  113 &  0.49 &  5.66 &   0.3 & $<$ 0.2 &  40.47 &    5.97 \\
       UM\,603   &  -12.52 &  -12.98 &   23 &  1.64 &  1.75 &  20.4 &  13.6   &  42.27 &    7.77 \\
Tol\,1358-328    &  -12.29 &  -12.75 &  166 &  0.37 &  4.27 &   0.1 & $<$ 0.1 &  40.01 &    5.51 \\
Tol\,2259-398    &  -12.01 &  -12.46 &   25 &  2.25 &  1.46 &  19.9 &   1.9   &  42.26 &    7.76 \\
Tol\,2306-400    &  -11.88 &  -12.33 &    7 &  3.71 &  1.49 & 135.6 &   7.6   &  43.09 &    8.59 \\
      MRK930     &  -12.03 &  -12.49 &   71 &  1.20 &  3.35 &   7.2 & $<$ 4.8 &  41.82 &    7.32 \\
\enddata
\tablenotetext{.}{Column 1: name of the galaxy; Columns 2 and 3: Logarithm of the
extinction corrected \Ha\ and \Hb\ emission  lines fluxes; Column 4: observed \Hb\ equivalent width;
Column 5: visual extinction; Column 6:[OIII]/H$\beta$; Columns 7: extiction corrected SFR(\Ha\ );
Column 8: SFR(1.4GHz); Column 9: Log L(\Ha ); Column 10: Mass of the burst from SB99 (see
text).}
\end{deluxetable}

\begin{deluxetable}{lrcrr}
\tablewidth{0pt}\tabletypesize{\scriptsize}
\tablecaption{\label{Tab:Tracers} Infrared fluxes, d and q parameters for our sample}
\tablehead{
\colhead{Name}      & \colhead{ $d$ }     & \colhead{F(60\,$\mu$m)}  
& \colhead{F(100\,$\mu$m)}  
& \colhead{$q$} \\
\colhead{ }    & \colhead{\% }       & \colhead{(Jy) }   & \colhead{(Jy)} & \colhead{\ }               
}
\startdata
       UM\,304       &    100  & 1.983    & 2.426       &       2.6  \\
Tol\,0127-397        & $<$ 19  & 0.233    & $<$0.351    & $>$   2.6  \\
Tol\,0145-391        & $<$  5  & --       & --          &       --   \\
Tol\,0226-390        &     42  & 0.791    & 0.480       &       2.4  \\
      MRK605         & $<$ 14  & 0.272    & $<$1.359    & $>$   2.6  \\
Tol\,0614-375        & $<$ 61  & 0.216    & $<$0.529    & $>$   2.0  \\
Tol\,0619-392        & $<$ 12  & 0.419    & $<$1.587    & $>$   2.8  \\
      MRK711         &     47  & 0.809    & 0.749       &       2.6  \\
Tol\,0957-278        &     70  & --       & --          &       --   \\
       MRK36         &     37  & 0.225    & $<$0.681    &       2.1  \\
Tol\,1116-325        & $<$ 75  & --       &  --         &       --   \\
       UM\,421       &     35  & 0.560    & $<$1.304    &       2.3  \\
       UM\,444       & $<$ 14  & --       & --          &       --   \\
       UM\,468       &     80  & 0.295    & 0.589       &       2.0  \\
       UM\,482       & $<$ 71  & --       & --          &       --   \\
     MRK1315         & $<$ 81  & --       & --          &       --   \\
     MRK1318         & $<$ 22  & 0.730    & 0.940       & $>$   2.8  \\
       UM\,488       &     40  & --       &  --         &       --   \\
       MRK52         &      5  & 4.430    & 6.650       &       2.7  \\
Tol\,1223-388        & $<$ 27  & 1.140    & 2.534       & $>$   3.1  \\
Tol\,1247-232        &     21  & 0.507    & $<$0.968    &       2.3  \\
       UM\,530       &     28  & 0.576    & 0.632       &       2.0  \\
       UM\,533       &     33  & 0.496    & 0.540       &       2.4  \\
Tol\,1303-281        & $<$ 54  & --       & --          &       --   \\
Tol\,1304-386        & $<$ 25  & --       & --          &       --   \\
Tol\,1324-276        &     56  & 0.557    & 0.740       &       2.3  \\
       UM\,603       &     67  & --       & --          &       --   \\
Tol\,1358-328        & $<$ 18  & 1.563    & 2.704       & $>$   3.1  \\
Tol\,2259-398        & $<$ 10  & --       &  --         &       --   \\
Tol\,2306-400        & $<$  6  & --       &  --         &       --   \\
      MRK930         &     67  &  1.245   & $<$2.154    &       2.2  \\
\enddata
\tablenotetext{.}{Column 1: Name of the galaxy; Column 2: $d$-parameter;
Columns 3 and 4: IRAS fluxes at 60\,$\mu$m and 100\,$\mu$m;
Column 5 $q$ parameter.}
\end{deluxetable}

\begin{deluxetable}{lrcrrrrr}
\tablewidth{0pt}\tabletypesize{\scriptsize}
\tablecaption{\label{Tab:Jansen} 
Star formation rate and d parameter of normal galaxies}
\tablehead{
\colhead{Name}          & \colhead{cz}                & \colhead{EW(H$\beta$)}  & \colhead{SFR(H$\alpha$)}  
& \colhead{F(1.4 GHz)}  & \colhead{SFR(1.4 GHz)}  & \colhead{$d$ parameter}  &\colhead{Ref.}\\
\colhead{\ }            & \colhead{(km/s)}              & \colhead{(\AA)}         & \colhead{(\Msolar year$^{-1}$)} 
& \colhead{(mJy)}         & \colhead{(\Msolar year$^{-1}$)} & \colhead{(\%)} &\colhead{}\\
}
\startdata
UGC00484  &  4859 &  4.3 & 2.7  &  4.4 & 1.5   & 56 &1\\
UGC01154  &  7756 &  6.0 & 5.8  &  5.6 & 5.0   & 88 &1\\
UGC01155  &  3158 &  7.5 & 0.7  &  4.4 & 0.6   & 92 &1\\
NGC695    &  9705 &  0.8 & 79.4 & 75.6 & 107.2 & 135 &1\\
UGC01551  &  2669 &  5.6 & 0.8  &  3.0 & 0.3   & 38 &1\\
IC197     &  6332 &  0.8 & 3.8  & 10.9 & 6.5   & 171 &1\\
UGC01630  &  4405 &  6.2 & 2.5  & 14.3 & 4.1   & 163 &1\\
NGC927    &  8258 &  1.0 & 10.2 & 16.7 & 17.0  & 166 &1\\
UGC04713  &  9036 &  0.8 & 19.5 &  2.9 & 3.5   & 18 &1\\
NGC2844   &  1486 &  1.7 & 0.1  &  3.9 & 0.1   & 89 &1\\
NGC3011   &  1517 &  4.2 & 0.05 &  3.0 & 0.1   & 200 &1\\
IC2520    &  1226 &  0.8 & 0.5  & 24.1 & 0.5   & 109 &1\\
NGC3075   &  3566 &  6.0 & 0.7  &  3.9 & 0.7   & 102 &1\\
NGC3104   &   604 & 10.4 & 0.04 &  1.8 & 0.01  & 23 &1\\
NGC3264   &   929 & 20.5 & 0.11 &  2.1 & 0.03  & 23 &1\\
NGC3279   &  1422 &  1.9 & 0.3  &  8.7 & 0.3   & 101 &1\\
UGC05744  &  3338 &  5.8 & 0.6  &  5.1 & 0.8   & 134 &1\\
UGC05760  &  2997 &  4.7 & 0.8  &  7.9 & 1.0   & 126 &1\\
IC2591    &  6755 &  1.1 & 2.7  &  5.3 & 3.6   & 134 &1\\
UGC05819  & 10693 &  0.3 & 2.3  &  5.4 & 9.3   & 402 &1\\
NGC3510   &   704 & 13.4 & 0.07 &  3.1 & 0.02  & 31 &1\\
IC673     &  3851 &  1.8 & 1.1  &  5.6 & 1.2   & 105 &1\\
NGC3633   &  2553 &  2.8 & 1.7  & 19.4 & 1.9   & 107 &1\\
UGC06625  & 10964 &  7.4 & 11.0 & 17.4 & 31.6  & 287 &1\\
IC746     &  5027 &  1.3 & 1.3  &  4.4 & 1.7   & 131 &1\\
NGC3978   &  9978 &  5.2 & 21.4 & 38.2 & 57.5  & 270 &1\\
UGC07020A &  1447 & 12.4 & 0.3  &  5.3 & 0.2   & 45 &1\\
NGC4120   &  2251 &  8.7 & 0.3  &  4.0 & 0.3   & 98 &1\\
NGC4159   &  1761 &  8.8 & 0.3  &  5.9 & 0.3   & 83 &1\\
NGC4238   &  2771 & 10.2 & 0.5  &  3.2 & 0.4   & 68 &1\\
UGC07358  &  3639 &  3.9 & 0.8  &  3.2 & 0.6   & 79 &1\\
UGC07690  &   540 &  7.6 & 0.02 &  1.6 & 0.01  & 33 &1\\
NGC4509   &   907 & 25.0 & 0.1  &  2.8 & 0.03  & 36 &1\\
UGC07761  &  6959 &  5.9 & 4.3  &  8.9 & 6.5   & 151 &1\\
NGC4758   &  1244 &  3.5 & 0.2  &  3.0 & 0.07  & 35 &1\\
NGC5230   &  6855 &  4.3 & 8.7  & 19.5 & 13.8  & 158 &1\\
UGC08630  &  2364 &  9.7 & 0.3  &  3.0 & 0.2   & 84 &1\\
NGC5425   &  2062 &  4.9 & 0.2  &  7.1 & 0.4   & 182 &1\\
NGC5491   &  5845 &  1.9 & 2.9  &  4.7 & 2.4   & 81 &1\\
NGC5541   &  7698 &  4.8 & 10.2 & 33.0 & 29.5  & 284 &1\\
UGC09356  &  2234 &  5.9 & 0.2  &  2.5 & 0.2   & 75 &1\\
UGC09560  &  1215 & 45.9 & 0.2  &  5.9 & 0.1   & 84 &1\\
IC1100    &  6561 &  0.9 & 3.6  &  7.6 & 4.9   & 134 &1\\
IC1124    &  5242 &  0.9 & 2.6  &  3.1 & 1.3   & 48 &1\\
IC1141    &  4458 &  0.3 & 1.3  & 10.7 & 3.2   & 230 &1\\
NGC6007   & 10548 &  3.3 & 7.8  & 13.5 & 22.9  & 292 &1\\
UGC10086  &  2191 & 10.3 & 0.3  &  3.9 & 0.3   & 108 &1\\
NGC6131   &  5054 &  4.2 & 1.0  &  6.5 & 2.5   & 247 &1\\
NGC7328   &  2827 &  1.9 & 1.4  &  6.9 & 0.8   & 58 &1\\
UGC12178  &  1925 &  3.6 & 0.4  &  7.2 & 0.4   & 96 &1\\
UGC12265  &  5682 & 12.9 & 3.2  & 14.1 & 6.8   & 208 &1\\
NGC7460   &  3296 &  5.9 & 2.6  & 13.3 & 2.1   & 81 &1\\
UGC12519  &  4380 &  6.2 & 2.2  &  4.6 & 1.3   & 57 &1\\
NGC7620   &  9565 &  7.1 & 31.6 & 31.5 & 43.7  & 139 &1\\
IC1504    &  6306 &  1.7 & 11.2 &  7.4 & 4.4   & 38 &1\\
NGC2782 &    2558 &  6.9 &  6.6 & 124.5&  12.2 & 185 & 2 \\
NGC3310 &     988 & 21.0 &  7.6 & 396.9&   9.8 & 128 & 2 \\
NGC3367 &    3041 &  6.2 &  9.7 & 117.6&  15.8 & 162 & 2 \\
NGC3432 &     615 & 11.1 &  0.5 &  83.3&   0.4 &  73 & 2 \\
NGC3504 &    1538 &  5.7 &  6.8 & 274.1&  13.6 & 199 & 2 \\
NGC3690 &    3033 & 21.8 & 33.3 & 686.3&  79.1 & 237 & 2 \\
NGC3893 &     971 &  5.7 &  4.0 & 139.2&   2.8 &  70 & 2 \\
NGC3949 &     796 &  7.2 &  2.5 & 118.4&   2.4 &  97 & 2 \\
NGC4088 &     759 &  5.9 &  4.5 & 222.0&   4.5 &  99 & 2 \\
NGC4102 &     837 &  2.3 &  2.7 & 272.7&   5.6 & 205 & 2 \\
NGC4157 &     775 &  3.5 &  2.1 & 190.0&   3.9 & 182 & 2 \\
NGC4214 &     291 & 20.9 &  0.2 &  37.2&   0.03&  14 & 2 \\
NGC4217 &    1026 &  1.3 &  1.1 & 120.3&   2.5 & 229 & 2 \\
NGC4254 &    2406 &  6.1 &  8.7 & 447.4&   8.9 & 102 & 2 \\
NGC4303 &    1569 &  6.3 &  5.3 & 444.3&   7.2 & 137 & 2 \\
NGC5676 &    2116 &  3.8 & 11.2 & 113.7&   9.6 &  84 & 2 \\
\enddata
\tablenotetext{.}{Column 1: Galaxy Name; Column 2: redshift;
Column 3: H$\beta$ equivalent width; Column 4: H$\alpha$ SFR;
Column 5: 1.4~GHz flux; Column 6: 1.4~GHz SFR; Column 7:
$d$-parameter; Column 8: reference from which the galaxy was
obtained (1- Jansen et al. 2000; 2- Ho et al. 1997)}
\end{deluxetable}

\begin{deluxetable}{lccrrccccc}
\tablewidth{0pt}\tabletypesize{\scriptsize}
\tablecaption{\label{Tab:Ho-Sample}
Radio and Infrared properties of galaxies from Ho et al. (1997)}
\tablehead{
\colhead{Name}      & \colhead{Type} & \colhead{Distance} 
& \colhead{F(1.4 GHz) }   & \colhead{F(4.8 GHz)} & \colhead{F(60 $\mu$m)}& 
\colhead{F(100 $\mu$m)}  & \colhead{$q$} &   \colhead{References} \\
\colhead{ }      & \colhead{ } & \colhead{ }             
& \colhead{(mJy) }   & \colhead{(mJy)} & \colhead{ (Jy)}& 
\colhead{ (Jy)}  & \colhead{ }  & \colhead{ } \\
}
\startdata
 NGC2782  & 1  &  37.3 &  124.5$\pm$5.1     &  49$\pm$   8 &     9.60  &   14.65  & 2.03    & 1\\
 NGC2964  & 4  &  21.9 &  105.2$\pm$3.9     &  33$\pm$   6 &    12.47  &   24.14  & 2.25    & 1\\
 NGC2967  & 5  &  30.9 &  111.5$\pm$3.7     &  49$\pm$  11 &     5.81  &   15.12  & 1.96    & 2\\
 NGC3310  & 4  &  18.7 &  396.9$\pm$12.5    & 152$\pm$  17 &    34.13  &   47.95  & 2.06    & 1\\
 NGC3367  & 5  &  43.6 &  117.6$\pm$4.2     &  36$\pm$   8 &     6.06  &   12.49  & 1.91    & 1\\
 NGC3432  & 9  &   7.8 &   83.3$\pm$3.4     &  40$\pm$   7 &     8.55  &   16.44  & 2.19    & 1\\
 NGC3504  & 2  &  26.5 &  274.1$\pm$10.6    & 117$\pm$  16 &    22.70  &   35.70  & 2.06    & 1\\
 NGC3556  & 6  &  14.1 &  284.1$\pm$8.0     &  76$\pm$   9 &    32.19  &   80.77  & 2.29    & 1\\
 NGC3583  & 3  &  34.0 &   59.2$\pm$2.4     &  30$\pm$   6 &     7.18  &   19.50  & 2.33    & 1\\
 NGC3593  & 0  &   5.5 &   86.2$\pm$3.4     &  55$\pm$  10 &    18.27  &   36.00  & 2.51    & 1\\
 NGC3631  & 5  &  21.6 &   84.0$\pm$3.5     &  31$\pm$   6 &     9.58  &   26.77  & 2.31    & 1\\
 NGC3690  & 9  &  40.4 &  686.3$\pm$26.5    & 300$\pm$   4 &   121.64  &  122.45  & 2.33    & 4\\
 NGC3810  & 5  &  16.9 &  124.3$\pm$4.7     &  46$\pm$   9 &    13.99  &   35.08  & 2.28    & 1\\
 NGC3893  & 5  &  17.0 &  139.2$\pm$5.0     &  39$\pm$   7 &    15.07  &   39.26  & 2.28    & 1\\
 NGC3949  & 4  &  17.0 &  118.4$\pm$4.2     &  37$\pm$   6 &    11.37  &   26.52  & 2.20    & 1\\
 NGC4088  & 4  &  17.0 &  222.0$\pm$6.3     &  67$\pm$   9 &    26.56  &   60.83  & 2.29    & 1\\
 NGC4102  & 3  &  17.0 &  272.7$\pm$9.6     &  70$\pm$   9 &    50.56  &   75.72  & 2.40    & 1\\
 NGC4157  & 3  &  17.0 &  190.0$\pm$6.4     &  60$\pm$   8 &    17.65  &   49.95  & 2.23    & 1\\
 NGC4214  &10  &   3.5 &   37.2$\pm$1.8     &  30$\pm$   7 &    17.87  &   29.04  & 2.83    & 1\\
 NGC4217  & 3  &  17.0 &  120.3$\pm$4.3     &  40$\pm$   7 &    11.70  &   41.40  & 2.30    & 1\\
 NGC4254  & 5  &  16.8 &  447.4$\pm$12.2    & 111$\pm$  16 &    44.00  &   96.32  & 2.20    & 1\\
 NGC4303  & 4  &  15.2 &  444.3$\pm$13.6    & 157$\pm$  23 &    41.00  &   77.40  & 2.14    & 1\\
 NGC4449  &10  &   3.0 &  264.0$\pm$7.2     & 108$\pm$  13 &    37.00  &   58.28  & 2.29    & 1\\
 NGC4490  & 7  &   7.8 &  807.9$\pm$20.8    & 226$\pm$  27 &    47.79  &   85.94  & 1.94    & 1\\
 NGC4532  &10  &  15.5 &  122.8$\pm$4.3     &  54$\pm$  10 &    10.00  &   14.88  & 2.05    & 1\\
 NGC4535  & 5  &  16.8 &   77.6$\pm$5.3     &  38$\pm$   9 &    14.00  &   31.82  & 2.47    & 1\\
 NGC4536  & 4  &  13.3 &  205.1$\pm$7.7     & 114$\pm$  12 &    28.66  &   44.63  & 2.29    & 1\\
 NGC4631  & 7  &   6.9 &  976.2$\pm$16.7    & 438$\pm$  66 &    82.90  &  208.66  & 2.16    & 3\\
 NGC4654  & 6  &  16.8 &  122.5$\pm$4.6     &  48$\pm$   9 &    14.70  &   34.40  & 2.30    & 1\\
 NGC4793  & 5  &  38.9 &  113.3$\pm$4.1     &  46$\pm$   8 &    12.88  &   29.28  & 2.27    & 1\\
 NGC5248  & 4  &  22.7 &  157.4$\pm$7.0     &  71$\pm$  12 &    20.71  &   49.08  & 2.34    & 1\\
 NGC5676  & 4  &  34.5 &  113.7$\pm$4.2     &  38$\pm$   6 &    12.45  &   31.55  & 2.27    & 1\\
 NGC5775  & 5  &  26.7 &  279.9$\pm$9.1     &  67$\pm$  12 &    23.41  &   51.35  & 2.13    & 1\\
 NGC5907  & 5  &  14.9 &  103.4$\pm$3.8     &  38$\pm$   6 &     8.78  &   45.76  & 2.35    & 1\\
 NGC5962  & 5  &  31.8 &   83.8$\pm$3.2     &  36$\pm$   8 &     8.99  &   20.79  & 2.25    & 1\\
 NGC6181  & 5  &  36.7 &   95.6$\pm$3.5     &  56$\pm$   9 &     9.35  &   21.00  & 2.20    & 1\\
\enddata
\tablenotetext{.}{The 1.4 GHz fluxes are from the NVSS catalogue and  
the references for the fluxes at 4.8 GHz are in column 8:
1- Gregory \& Condon (1991), 
2- Griffith, M. R., et al. (1995),
3- Becker, White \& Edwards (1991) and 
4- Condon, Anderson \& Broderick (1995).
}
\end{deluxetable}

\begin{deluxetable}{lccccc}
\tablewidth{0pt}\tabletypesize{\scriptsize}
\tablecaption{\label{Tab:Summary} Tracers of a 
very young star forming episode}
\tablehead{
\colhead{Name}      & \colhead{High EW(\Hb)} & \colhead{High OIII/\Hb} 
& \colhead{High $\alpha^{1.4}_5$ }   & \colhead{Low $d$} & \colhead{High $q$}
 \\
\colhead{\ } & \colhead{$\geqslant 70$ \AA} & \colhead{$\geqslant 2$} 
&\colhead{$>-0.5$} & \colhead{$ \leqslant75$ \%} &\colhead{$> 2.5$} 
}
\startdata
       UM\,304   &   X     & X       & $\surd$  &  X          & $\surd$ \\
Tol\,0127-397    &   X     & $\surd$ & X        &  $\surd$    & $\surd$ \\
Tol\,0145-391    &   X     & X       & --       &  $\surd$    & --      \\
Tol\,0226-390    & $\surd$ & $\surd$ & $\surd$  &  $\surd$    & X       \\
      MRK605     &   X     & X       & $\surd$  &  $\surd$    & $\surd$ \\
Tol\,0614-375    &   X     & X       & --       &  $\surd$    & ?       \\
Tol\,0619-392    &   X     & X       & $\surd$  &  $\surd$    & $\surd$ \\
      MRK711     &   X     & $\surd$ & $\surd$  &  $\surd$    & $\surd$ \\
Tol\,0957-278    &   X     & $\surd$ & --       &  $\surd$    & --      \\
       MRK36     & $\surd$ & $\surd$ & --       &  $\surd$    & X       \\
Tol\,1116-325    & $\surd$ & $\surd$ & --       &  $\surd$    & --      \\
       UM\,421   &   X     & X       & $\surd$  &  $\surd$    & X       \\
       UM\,444   &   X     & $\surd$ & $\surd$  &  $\surd$    & --      \\
       UM\,468   &   X     & $\surd$ &  X       &  X          & X       \\
       UM\,482   &   X     & $\surd$ & --       &  $\surd$    & --      \\
     MRK1315     & $\surd$ & $\surd$ & --       &  ?          & --      \\
     MRK1318     & $\surd$ & $\surd$ & $\surd$  &  $\surd$    & $\surd$ \\
       UM\,488   &   X     & $\surd$ &  X       &  $\surd$    & --      \\
       MRK52     &   X     & X       & --       &  $\surd$    & $\surd$ \\
Tol\,1223-388    &   X     & X       & $\surd$  &  $\surd$    & $\surd$ \\
Tol\,1247-232    & $\surd$ & $\surd$ & $\surd$  &  $\surd$    & X       \\
       UM\,530   &   X     & $\surd$ &  X       &  $\surd$    & X       \\
       UM\,533   & $\surd$ & $\surd$ & $\surd$  &  $\surd$    & X       \\
Tol\,1303-281    & $\surd$ & $\surd$ & --       &  $\surd$    & --      \\
Tol\,1304-386    & $\surd$ & $\surd$ & $\surd$  &  $\surd$    & --      \\
Tol\,1324-276    & $\surd$ & $\surd$ & --       &  $\surd$    & X       \\
       UM\,603   &   X     & X       & X        &  $\surd$    & --      \\
Tol\,1358-328    & $\surd$ & $\surd$ & $\surd$  &  $\surd$    & $\surd$ \\
Tol\,2259-398    &   X     & X       & $\surd$  &  $\surd$    & --      \\
Tol\,2306-400    &   X     & X       & $\surd$  &  $\surd$    & --      \\
      MRK930     & $\surd$ & $\surd$ & --       &  $\surd$    & X       \\
\enddata
\end{deluxetable}

\end{document}